\documentclass[%
 reprint,
 amsmath,amssymb,
 aps,
]{revtex4-2}

\usepackage{graphicx}
\usepackage{dcolumn}
\usepackage{bm}
\usepackage{hyperref}
\usepackage{slashed}
\usepackage{multirow}
\usepackage{cancel}
\usepackage[bbgreekl]{mathbbol}
\usepackage{mathrsfs}  
\usepackage{amsmath}
\usepackage{relsize}
\usepackage{braket}
\usepackage{amsmath}
\usepackage{comment}
\usepackage{makecell}
\usepackage{tikz,xcolor}
\usepackage{tikzsymbols}

\newcommand{\aBL}{\alpha_{\mathsmaller{\rm B-L}}}
\newcommand{\chk}{\checkmark}
\newcommand{\gBL }{g_{\mathsmaller{\rm B-L}}}
\newcommand{\hc}{{\rm h.c.}}

\newcommand{\mX }{m_{\mathsmaller{\rm X}}}
\newcommand{\nn}{\nonumber}
\newcommand{\lX }{\lambda^{\mathsmaller{\rm X}}}
\newcommand{\qBL}{q_{\mathsmaller{\rm B-L}}}

\newcommand{\vBL}{v_{\mathsmaller{\rm B-L}}}
\newcommand{\vEW}{v_{\mathsmaller{\rm EW}}}
\newcommand{\vrel}{v_{\rm rel}}
\newcommand{\vMol}{v_{\text{M\o l}}}

\newcommand{\y}{\mathrm{y}}
\newcommand{\w}{\mathrm{w}}

\newcommand{\tikzxmark}{%
\tikz[scale=0.23] {
    \draw[line width=0.7,line cap=round] (0,0) to [bend left=6] (1,1);
    \draw[line width=0.7,line cap=round] (0.2,0.95) to [bend right=3] (0.8,0.05);
}}
\usepackage{tikz}
\usetikzlibrary{arrows,shapes}
\usetikzlibrary{trees,patterns}
\usetikzlibrary{matrix,arrows} 				
\usetikzlibrary{positioning}				  
\usetikzlibrary{calc,through}				  
\usetikzlibrary{decorations.pathreplacing}  
\usepackage{pgffor}							

\usetikzlibrary{decorations.pathmorphing}	
\usetikzlibrary{decorations.markings}
\tikzset{
	>=stealth', 
    vector/.style={decorate, decoration={snake}, draw},
	provector/.style={decorate, decoration={snake,amplitude=2.5pt}, draw},
	antivector/.style={decorate, decoration={snake,amplitude=-2.5pt}, draw},
	bigvector/.style={decorate, decoration={snake,amplitude=4pt}, draw},
    fermion/.style={draw=black, postaction={decorate},
        decoration={markings,mark=at position .55 with {\arrow[draw=black]{>}}}},
    fermionbar/.style={draw=black, postaction={decorate},
        decoration={markings,mark=at position .55 with {\arrow[draw=black]{<}}}},
    fermionnoarrow/.style={draw=black},
    gluon/.style={decorate, draw=black,
        decoration={coil,amplitude=4pt, segment length=5pt}},
    scalar/.style={dashed,draw=black, postaction={decorate},
        decoration={markings,mark=at position .55 with {\arrow[draw=black]{>}}}},
    scalarbar/.style={dashed,draw=black, postaction={decorate},
        decoration={markings,mark=at position .55 with {\arrow[draw=black]{<}}}},
    scalarnoarrow/.style={dashed,draw=black},
    momentum/.style={draw=black, postaction={decorate},
        decoration={markings,mark=at position 1 with {\arrow[draw=black]{>}}}},
    antimomentum/.style={draw=black, postaction={decorate},
        decoration={markings,mark=at position 0.1 with {\arrow[draw=black]{<}}}}
}

\tikzstyle{block} = [draw, rectangle, 
    minimum height=3em, minimum width=6em]

\DeclareMathOperator*{\SumInt}{%
\mathchoice%
  {\ooalign{$\displaystyle\sum$\cr\hidewidth$\displaystyle\int$\hidewidth\cr}}
  {\ooalign{\raisebox{.14\height}{\scalebox{.7}{$\textstyle\sum$}}\cr\hidewidth$\textstyle\int$\hidewidth\cr}}
  {\ooalign{\raisebox{.2\height}{\scalebox{.6}{$\scriptstyle\sum$}}\cr$\scriptstyle\int$\cr}}
  {\ooalign{\raisebox{.2\height}{\scalebox{.6}{$\scriptstyle\sum$}}\cr$\scriptstyle\int$\cr}}
}
\usepackage{bbm} 
\def\im	{\mathbbm{i}} 
\usepackage[capitalise]{cleveref}
\crefformat{pluralequation}{#2{\color{black} Eqs.~(}#1{\color{black} )}#3}
\Crefformat{pluralequation}
{#2{\color{black} Equations~(}#1{\color{black} )}#3}

\hypersetup{
	colorlinks=true,
	linkcolor=[rgb]{0.773,0.09,0.1},
	citecolor=blue,
	filecolor=magenta,      
	urlcolor=blue
}

\begin{document}

\preprint{APS/123-QED}

\title{
Critical and super-critical scatterings in baryogenesis and leptogenesis 
}

\author{Marcos M. Flores}
\email{marcos.flores@phys.ens.fr}
\author{Kalliopi Petraki}%
\email{kalliopi.petraki@phys.ens.fr}
\author{Anna Socha}
\email{anna.socha@phys.ens.fr}
\affiliation{Laboratoire de Physique de l'\'{E}cole normale sup\'{e}rieure, ENS, Université PSL, CNRS, Sorbonne Universit\'{e}, Universit\'{e} Paris Cit\'{e}, F-75005 Paris, France}%
\date{\today}

\begin{abstract}
In many theories, matter–antimatter asymmetries originate from out-of-equilibrium decays and scatterings of heavy particles. While decays remain efficient, scattering rates typically drop below the Hubble rate as the universe expands. We point out the possibility of scatterings between non-relativistic particles and the relativistic bath whose cross-sections grow with decreasing temperature, leading to scattering rates that track or exceed the Hubble rate at late times. This results in soaring asymmetry generation, even at low scales and with small CP- or baryon/lepton-violating couplings.

\end{abstract}

\maketitle


\textbf{\label{sec:Intro}Introduction \textemdash }
The origin of atoms, stars and galaxies can be traced back to a minuscule excess of baryons over antibaryons in the early universe. The asymmetry between matter and antimatter, quantified by the baryon-to-entropy density ratio, $Y_{\rm B} \equiv (n_{\rm B} - n_{\bar {\rm B}})/s = (8.69\pm 0.01)\times 10^{-11}$~\cite{Planck:2018vyg}, has been independently inferred from the primordial abundance of light nuclei~\cite{Cooke:2017cwo, Riemer-Sorensen:2017vxj} and the cosmic microwave background anisotropies. 
Generating an asymmetry requires processes satisfying the well-known Sakharov conditions~\cite{Sakharov:1967dj}. The Standard Model (SM) fails to do so sufficiently, making the observed baryon asymmetry a strong indication of new physics. 
It is also plausible that the dark matter (DM) density is due to a dark asymmetry, potentially linked to $Y_{\rm B}$~\cite{Petraki:2013wwa,Zurek:2013wia}. 

In some of the most well-motivated and testable scenarios, the matter-antimatter asymmetry arises from out-of-equilibrium, CP-violating decays~\cite{Weinberg:1979bt, Fry:1980bd, Fukugita:1986hr} or scatterings~\cite{Yoshimura:1978ex, Kolb:1979qa, Bento:2001rc} of heavy particles. These include leptogenesis~\cite{Fukugita:1986hr, Buchmuller:2005eh, Davidson:2008bu}, GUT baryogenesis~\cite{Yoshimura:1978ex, Kolb:1979qa, Langacker:1980js}, and analogous mechanisms for asymmetric DM~\cite{Falkowski:2011xh,Davidson:2012fn}. Decay processes become efficient, and remain so, once the Hubble parameter drops below the decay rate of the heavy species. In contrast, the scattering rates, $\mathcal{R} = \langle \sigma \vMol \rangle n_T$, are governed by the temperature dependence of the thermally-averaged cross-sections, which we may parametrize as $\langle \sigma \vMol \rangle \propto  T^n/\Lambda^{n+2}$ with $\Lambda$ being some energy scale, as well as the number density of the target particles, which, for relativistic species, scales as $n_T \sim T^3$. In existing models, $n > -1$, so scattering processes become inefficient as their rates decrease faster than the Hubble parameter during radiation domination.

In this \emph{letter}, we identify a novel possibility, where particle-number-violating and/or CP-violating scatterings with $n \leqslant -1$ drive asymmetry generation. For $n=-1$, the scattering rate scales with the Hubble parameter and exceeds it for sufficiently large couplings. More strikingly, for $n<-1$, the scattering rate inevitably overtakes the Hubble expansion at low temperatures, regardless of the coupling strength. We refer to these scalings as \emph{critical} and \emph{super-critical}, respectively.  Notably, these dynamics can lead to highly efficient asymmetry generation, even at lower energy scales, potentially within the reach of future experiments.

We consider, specifically, a two-flavor scenario where the CP-violating scatterings of heavy fermions $X$ and relativistic scalars $\rho$ exhibit \emph{super-critical} scaling, leading to prodigious asymmetry generation. The asymmetry eventually freezes out as the heavier fermion population is depleted through $X$-violating decays and \emph{critical} scatterings with the thermal bath. As we shall see, this behavior stems directly from the charge assignments of the interacting fields under a $U(1)_{\rm B-L}$ gauge symmetry.

Such model structures are broadly motivated: scalar mediators commonly appear in asymmetry-generation scenarios, where their interactions provide the necessary CP violation. In compelling models, scalars typically carry gauge charges that allow them to play a role in spontaneous symmetry breaking. These features naturally give rise to interactions exhibiting \emph{critical} or \emph{super-critical} scaling, suggesting that the dynamics identified here may be generic in theories of asymmetry generation.

\emph{Critical} and \emph{super-critical} scatterings in the early universe have previously been shown to strongly affect DM production~\cite{Binder:2023ckj}. In that case, \emph{(super-)critical} annihilations of two non-relativistic DM particles prevent DM chemical decoupling, but this behavior is ultimately unphysical: \emph{super-criticality} arises for $\sigma \propto 1/k_{\mathsmaller{\rm CM}}^\gamma$ with $\gamma \geqslant 3$~\cite{Binder:2023ckj}, where $k_{\mathsmaller{\rm CM}}$ is the momentum in the center-of-momentum frame; this 
violates unitarity at low $k_{\mathsmaller{\rm CM}}$~\cite{Flores:2024sfy}. Proper unitarization curtails this behavior and ensures freeze-out~\cite{Petraki:2025zvv}. In contrast, for scatterings on relativistic targets, as considered here, \emph{critical} behavior occurs already for $\gamma \geqslant 1$, consistent with unitarity at low energies.
\\

\textbf{The model \textemdash  } We extend the SM gauge group by a  $U(1)_{\rm B-L}$ factor, under which the SM quarks and leptons have the standard charges, $1/3$ and $-1$, respectively. Besides the $B-L$ gauge boson $V_\mu$, we introduce a complex scalar $\rho$, and two Dirac fermions, $X_1$ and $X_2$. The $\rho$ and $X_{1,2}$ fields are SM singlets, with $B-L$ charges $-2$ and $-1$,
respectively. Their interactions generate the desired asymmetry that cascades down to SM fermions, as described below. To cancel the gauge and gravitational anomalies, we also introduce three SM-singlet right-handed fermions $\chi$, and discern two scenarios: 
\\[1ex]
\textit{Type A:} All $\chi$ fermions have $B-L$ charge $-1$. They are stabilized by a $\mathbb{Z}_2$ symmetry, under which they are odd, while all other particles are even.
\\[1ex]
\textit{Type B:}  
The $\chi$ fermions have $B-L$ charges $\{5, -4, -4\}$. 
This assignment cancels the anomalies, while precluding renormalizable couplings to other fields. 
\\[1ex]
The $\chi$ particles play no role in asymmetry generation, we thus  discuss the two scenarios further in \cref{app:Cosmo}.

With these assignments, the Lagrangian density reads
\begin{align}
\mathcal{L} 
= \mathcal{L}_{\rm SM} 
+ \mathcal{L}_{\rm BSM}^0 
+ \mathcal{L}_{\rm BSM}^{\rm int}
- V_{\rm BSM}^{\rho}  , 
\label{eq:L}
\end{align}
where $\mathcal{L}_{\rm SM}$ denotes the SM Lagrangian.  $\mathcal{L}_{\rm BSM}^0$ contains the kinetic terms of the new fields and the $X$ mass terms,
\begin{align}
\mathcal{L}_{\rm BSM}^0 
&\equiv 
- \frac{1}{4} F_{\mu \nu} F^{\mu \nu} 
+ (\mathcal{D}_\mu \rho)^\dagger (\mathcal{D}^\mu \rho) 
+ \frac{1}{2} \sum_{k=1}^3 \bar{\chi}_{k}\im  \slashed{\mathcal{D}} \chi_{k} 
\nn \\
&+  \sum_{i=1}^2 \left(\bar{X}_i \im \slashed{\mathcal{D}} X_i - m_i \bar{X}_i X_i \right), 
\label{eq:LBSM_0}
\end{align}
where $F_{\mu \nu} = \partial_\mu V_\nu - \partial_\nu V_\mu$, and $\mathcal{D}_\mu = \mathcal{D}_\mu^{\rm SM} + 
\im \gBL \qBL  V_\mu$ 
with $\gBL$ being the $B-L$ gauge coupling, and $\aBL\equiv \gBL^2/(4\pi)$. We have diagonalized the $X$ mass matrix, and take $\Delta \mX \equiv m_2 - m_1 >0$, while $\mX$ will refer collectively to $m_1$ and $m_2$; we will generally assume that the two masses are similar, $\Delta \mX/\mX \lesssim {\cal O}(0.1)$. The interaction Lagrangian is 
\begin{align}
\label{eq:IntLag}
\mathcal{L}_{\rm BSM}^{\rm int} \equiv 
&- \sum_{i,j =1}^2  \left(\frac{y_{ij}^{\rm R}}{2^{\delta_{ij}}}  \rho^\dagger \Bar{X}^c_{\mathrm{R},i} X_{\mathrm{R},j} + \frac{y_{ij}^{\rm L}}{2^{\delta_{ij}}}  \rho^\dagger \Bar{X}^c_{\mathrm{L},i} X_{\mathrm{L},j}  
\right)
\nn \\
&-\sum_{l=1}^3 \sum_{i=1}^2
\lX_{il} \, \Bar{L}_l \tilde{\mathcal{H}}  X_{\mathrm{ R},i} 
+ \hc,
\end{align}
with $L_l$ and $\mathcal{H}$ being respectively the SM left-handed lepton and the Higgs doublet,
 $\Tilde{\mathcal{H}} \equiv \im  \sigma^2 \mathcal{H}^*$, and $X_{\mathrm{L} (\mathrm{R}), i} \equiv (1 \mp \gamma^5) X_i/2$. 
The Yukawa matrices $y^{\rm R, L}$ are generally complex, with $N^2+N$ phases for $N$ families of $X$ fermions. $X_{\rm {L,R}}$ rotations remove the $2N$ phases, $N$ of which render the $X$ masses real. Rotating $\rho$ amounts to rotating all $X_{\rm {L,R}}$ by half, due to the global $B-L$ symmetry, and does not remove any phase. The number of irreducible phases is thus $N^2=4$, for $N=2$ generations. For convenience, we define
\begin{align}
&\y_{ij} \equiv  \frac{y_{ij}^{\rm L} +  y_{ij}^{\rm R}}{2}, \quad
&\w_{ij} \equiv  \frac{y_{ij}^{\rm L} -  y_{ij}^{\rm R}}{2}, && i,j \in \{1,2\}.
\label{eq:Yukawas_redef}
\end{align}
The $X_{\rm {L,R}}$ rephasings can make $\y_{11}$ and $\y_{22}$ real, but cannot remove the $\w_{ij}$ phases without reintroducing complex phases to the $X$ masses. 
Finally, the scalar potential is
\begin{align}
V_{\rm BSM}^{\rho } \equiv 
- \mu_\rho^2 |\rho|^2 
+  \frac{1}{4}\lambda_\rho |\rho|^4 
+ \lambda_{\rho {\cal H}} |{\cal H}|^2 |\rho|^2,
\label{eq:Vrho}
\end{align}
where $\mu_\rho^2>0$ implies $B-L$ breaking at low temperatures, while $\lambda_\rho > 0$ and $\lambda_{\rho {\cal H}} > - 2 \sqrt{\lambda_\rho \lambda_{\cal H}}$ ensure vacuum stability and $\lambda_{\cal H}$ is the quartic Higgs coupling~\cite{Kannike:2012pe, Chakrabortty:2013mha}. \\

\textbf{Cosmology \textemdash  } We discern three major energy scales.
\\[1ex]
\textit{High energies:} 
Well above the TeV scale, the $B-L$ and electroweak (EW) symmetries are unbroken. We assume that the $X$ fermions have masses much larger than both breaking scales (indicatively, $\mX\sim$~10~PeV). 
Although thermal effects endow $\rho$ particles with temperature-dependent masses, $m_\rho(T)$ (see \cref{eq:m_rho}), they remain relativistic during this stage. At high temperatures, the $X$ fermions are kept in equilibrium via $X\bar{X} \leftrightarrow \rho \rho^*, V_\mu V_\mu$, generated by $y_{ij}^{\rm L,R}$ and $\gBL$, while the much smaller $\lX$ couplings to SM particles are ineffective at this stage. The $X$ particles begin to freeze-out at $T \sim \mX / 10$. The decays $X_2 \rightarrow \rho \bar{X}_1$, and a variety of $X-$ and CP-violating scatterings (cf.~\cref{App:CrossSections}) sequester global charge in $X$ and $\rho$, such that $\Delta_X \equiv \Delta_1 + \Delta_2  =-2 \Delta_\rho \neq 0$, where $\Delta_j \equiv Y_j - Y_{\bar{j}}$ for $j=X_1, X_2, \rho$. 
\\[1em]
\textit{Intermediate energies:} Before the $B-L$ and EW phase transitions, the $X_1 \to L_l {\cal H}$ decays transfer the $X$ global charge to the visible sector. The lepton asymmetry is converted into a baryon asymmetry via sphalerons~\cite{Rubakov:1996vz}. 
\\[1em]
\textit{Low energies:}
At $T = T_{\cancel{B-L}}$ ($\sim$~10~TeV, indicatively), $\rho$ develops a vacuum expectation value, breaking $U(1)_{\rm B-L}$. The radial $\rho$ component and $V_\mu$ acquire masses. Their densities are depleted via ($B-L$)-violating decays into SM, erasing $\Delta_\rho$ and inducing a net $B-L$ charge.
\\

Next, we discuss the CP-violating and $X$-violating processes in more detail, often referring to them according to the $B-L$ charge of their initial and final states. 
\\

\textbf{CP violation via super-critical scatterings~\textemdash  } 
The two-flavor model considered here does not allow for CP violation in any of the $X$-violating processes. CP violation arises however in $X$-conserving but flavor-changing 2-to-2 interactions. Intriguingly, our scenario features CP violation i) \emph{perturbatively}, through the interference of tree and one-loop diagrams, and ii) \emph{non-perturbatively}, in long-range $X_i \bar{X}_j \leftrightarrow X_{i'} \bar{X}_{j'}$ and $X_i \bar{X}_j \leftrightarrow \rho \rho^*$ interactions, via the resummation of one-boson exchange diagrams. Here, we concentrate on \emph{perturbative} CP violation, which, as we shall demonstrate, gives rise to \emph{super-critical asymmetry generation}. The non-perturbative CP violation merits dedicated discussion that we defer to a companion paper~\cite{Flores:2025b}. 

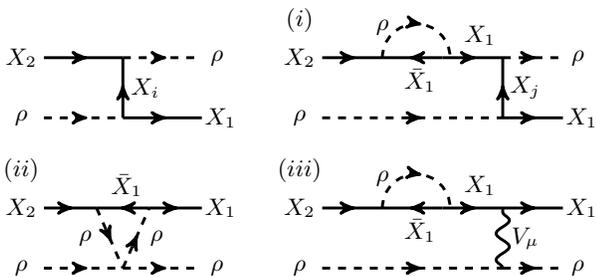
\begin{figure}[t!]
\centering
\begin{tikzpicture}[line width=1.1pt, scale=1]
\begin{scope}
\node at (-2,.8) {$X_2$};
\node at (-2,0) {$\rho$};
\node at (0.6,.8) {$\rho$};
\node at (-0.35,0.4) {$X_i$};
\node at (0.65,0) {$X_{1}$};
\draw[fermion] (-.65,0) -- (-.65,.8);
\draw[fermion]  (-1.7,.8) -- (-.65,.8);
\draw[scalar] (-1.7,0) -- (-.65,0);
\draw[scalar] (-.65,.8) -- (.4,.8);
\draw[fermion] (-.65,0) -- (.4,0);
\end{scope}
\begin{scope}[shift={(0,-2)}]
\node at (-2, 1.3) {$(ii)$};
\node at (-2,.8) {$X_2$};
\node at (-2,0) {$\rho$};
\node at (0.65,.8) {$X_1$};
\node at (0.6,0) {$\rho$};
\node at (-0.6,1.1) {$\bar{X}_1$};
\node at (-1.15,.4) {$\rho$};
\node at (-0.2,.4) {$\rho$};
\draw[scalar] (-1,.8) -- (-.65,0);
\draw[scalar] (-.65,0) -- (-.3,.8);
\draw[fermion]  (-1.7,.8) -- (-1,.8);
\draw[scalar] (-1.7,0) -- (-.65,0);
\draw[fermion] (-.3,.8) -- (-1,.8);
\draw[fermion] (-.3,.8) -- (.4,.8);
\draw[scalar] (-.65,0) -- (.4,0);
\end{scope}
\begin{scope}[shift={(5.5,0)}]
\node at (-3.8, .8) {$X_2$};
\node at (-3.8, 1.3) {$(i)$};
\node at (-3.8,0) {$\rho$};
\node at (-0.1,0.8) {$\rho$};
\node at (-0.05,0) {$X_{1}$};
\node at (-0.8,0.4) {$X_j$};
\node at (-2.7,1.2) {$\rho$};
\node at (-2.2,.5) {$\bar{X}_1$};
\node at (-1.4,1.1) {$X_1$};
\draw[fermion] (-3.5,.8) -- ( -2.7,.8);
\draw[scalar] (-2.7,.8) arc (180:0:.45);
\draw[fermion] (-1.9,.8) -- ( -2.7,.8);
\draw[fermion] (-1.9,.8) -- ( -1.1,.8);
\draw[fermion] (-1.1,0) -- (-1.1,.8);
\draw[scalar] (-3.5,0) -- (-1.1,0);
\draw[scalar] (-1.1,.8)--(-.3,.8);
\draw[fermion] (-1.1,0)-- (-.3,0);
\end{scope}
\begin{scope}[shift={(+5.5,-2)}]
\node at (-3.8, 1.3) {$(iii)$};
\node at (-3.8, .8) {$X_2$};
\node at (-3.8,0) {$\rho$};
\node at (-0.05,0.8) {$X_1$};
\node at (-.1,0) {$\rho$};
\node at (-0.8,0.4) {$V_\mu$};
\node at (-2.7,1.2) {$\rho$};
\node at (-2.2,.5) {$\bar{X}_1$};
\node at (-1.4,1.1) {$X_1$};
\draw[fermion] (-3.5,.8) -- ( -2.7,.8);
\draw[scalar] (-2.7,.8) arc (180:0:.45);
\draw[fermion] (-1.9,.8) -- ( -2.7,.8);
\draw[fermion] (-1.9,.8) -- ( -1.1,.8);
\draw[vector] (-1.1,0) -- (-1.1,.8);
\draw[scalar] (-3.5,0) -- (-1.1,0);
\draw[fermion] (-1.1,.8)--(-.3,.8);
\draw[scalar] (-1.1,0)-- (-.3,0);
\end{scope}
\end{tikzpicture}
\caption{Diagrams contributing to the CP asymmetry for the $\qBL=-3$ states. The chosen flavor assignment ensures the presence of a branch cut, corresponding to on-shell intermediate states. The arrows denote the $-(B-L)$ charge flow. Loop diagram $(ii)$ is subdominant and we neglect it.}
\label{fig:CPdiag3}
\end{figure}

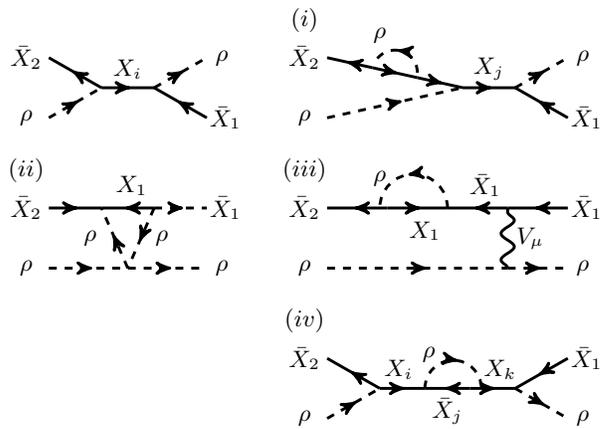
\begin{figure}[t!]
\centering
\begin{tikzpicture}[line width=1.1pt, scale=1]
\begin{scope}
\node at (-2,.8) {$\bar{X}_2$};
\node at (-2,0) {$\rho$};
\node at (0.6,.8) {$\rho$};
\node at (-0.65,0.65) {$X_i$};
\node at (0.65,0) {$\bar{X}_{1}$};
\draw[fermion] (-1,.4) -- (-.3,0.4);
\draw[fermion] (-1,.4)-- (-1.7,.8);
\draw[scalar] (-1.7,0) -- (-1,0.4);
\draw[scalar] (-.3,.4) -- (.4,.8);
\draw[fermion] (.4,0) -- (-.3,0.4) ;
\end{scope}
\begin{scope}[shift={(0,-2)}]
\node at (-2,1.3) {$(ii)$};
\node at (-2,.8) {$\bar{X}_2$};
\node at (-2,0) {$\rho$};
\node at (0.65,.8) {$\bar{X}_1$};
\node at (0.6,0) {$\rho$};
\node at (-0.6,1.1) {$X_1$};
\node at (-1.15,.4) {$\rho$};
\node at (-0.2,.4) {$\rho$};
\draw[scalar]  (-.65,0) -- (-1,.8);
\draw[scalar]  (-.3,.8) -- (-.65,0);
\draw[fermion]  (-1.7,.8) -- (-1,.8);
\draw[scalar] (-1.7,0) -- (-.65,0);
\draw[fermion] (-.3,.8) -- (-1,.8);
\draw[scalar] (-.3,.8) -- (.4,.8);
\draw[scalar] (-.65,0) -- (.4,0);
\end{scope}
\begin{scope}[shift={(5.5,0)}]
\node at (-3.8, 1.3) {$(i)$};
\node at (-3.8, .8) {$\bar{X}_2$};
\node at (-1.35, .65) {$X_j$};
\node at (-3.8,0) {$\rho$};
\node at (-0.1,0.8) {$\rho$};
\node at (-0.05,0) {$\bar{X}_{1}$};
\node at (-2.8,1.1) {$\rho$};
\draw[fermion]  ( -2.9,.8-0.133) -- (-3.5,.8);
\draw[scalar] (-2.3,.8-0.2) arc (-180:-30:-.3);
\draw[fermion]  ( -2.9,.8-0.133) -- (-2.3,.8-0.133-0.133);
\draw[fermion]  ( -2.3,.8-0.133-0.133) -- (-1.7,.8-0.133-0.133-0.133);
\draw[fermion]  (-1.7,.4) -- (-1,.4);
\draw[scalar] (-3.5,0) -- (-1.7,0.4);
\draw[scalar] (-1,.4)--(-.3,.8);
\draw[fermion]  (-.3,0) -- (-1,0.4);
\end{scope}
\begin{scope}[shift={(+5.5,-2)}]
\node at (-3.8, 1.3) {$(iii)$};
\node at (-3.8, .8) {$\bar{X}_2$};
\node at (-3.8,0) {$\rho$};
\node at (-0.05,0.8) {$\bar{X}_1$};
\node at (-.1,0) {$\rho$};
\node at (-0.8,0.4) {$V_\mu$};
\node at (-2.8,1.2) {$\rho$};
\node at (-2.2,.5) {$X_1$};
\node at (-1.4,1.1) {$\bar{X}_1$};
\draw[fermion]( -2.7,.8) --  (-3.5,.8);
\draw[scalar] (-1.9,.8) arc (-180:0:-.45);
\draw[fermion]  ( -2.7,.8) -- (-1.9,.8);
\draw[fermion]  ( -1.1,.8) -- (-1.9,.8);
\draw[vector] (-1.1,0) -- (-1.1,.8);
\draw[scalar] (-3.5,0) -- (-1.1,0);
\draw[fermion] (-.3,.8) -- (-1.1,.8);
\draw[scalar] (-1.1,0)-- (-.3,0);
\end{scope}
\begin{scope}[shift={(5.5,-4)}]
\node at (-3.8, 1.3) {$(iv)$};
\node at (-3.8,.8) {$\bar{X}_2$};
\node at (-3.8,0) {$\rho$};
\node at (-.05,.8) {$\bar{X}_1$}; 
\node at (-.1,0) {$\rho$};
\node at (-2.55,.65) {$X_i$};
\node at (-1.2,.65) {$X_k$};
\node at (-1.9,.1) {$\bar{X}_j$};
\node at (-2.15,.85) {$\rho$};
\draw[fermion]  ( -2.8,.4)--(-3.5,.8);
\draw[scalar] (-3.5,0) -- (-2.8,0.4);
\draw[scalar] ( -2.2,.4) arc (180:0:.375);
\draw[fermion]  ( -2.8,.4)--( -2.2,.4);
\draw[fermion]  ( -1.6,.4)--( -2.2,.4);
\draw[fermion]  ( -1.6,.4)--(-1,.4);
\draw[fermion] (-0.3,.8) -- (-1,.4);
\draw[scalar] (-1,0.4)--(-.3,0);
\end{scope}
\end{tikzpicture}
\caption{As in \cref{fig:CPdiag3}, for the $\qBL=-1$ states.} 
\label{fig:CPdiag1}
\end{figure}

Perturbative CP violation occurs in
$X_i X_j\leftrightarrow X_{i'}X_{j'}$ and $X_i\rho^{(*)}\leftrightarrow X_{j'}\rho^{(*)}$ scatterings. The former experience significant Sommerfeld suppression, rendering the charge-3 and charge-1 flavor-changing interactions the dominant sources of CP violation. The relevant diagrams are shown in \cref{fig:CPdiag3,fig:CPdiag1}, while the corresponding CP asymmetry coefficients, $\varepsilon$, are calculated using the Cutkosky rules \cite{Cutkosky:1960sp} and collected in \cref{tab:CrossSections}. 

While the $X_i \rho^{(*)} \rightarrow X_j \rho^{(*)}$ processes resemble Thomson scattering, they differ from it in two important ways: (a) the fermion-fermion-scalar vertices are \emph{not} momentum-suppressed for non-relativistic $X$, and (b) at tree-level, they occur \emph{only} via $u$-channel \emph{or} $s$-channel exchange of $X_i$ fermions, as shown in \cref{fig:CPdiag3,fig:CPdiag1}. Each of the $s$- and $u$-channel diagrams contributes to the amplitude by $\delta{\cal M} \propto 1/k_{\mathsmaller{\rm CM}}$, at leading order in $k_{\rm CM} \ll \mX$. If $\rho$ were uncharged, both diagrams would contribute to each process and interfere destructively, with their leading-order terms canceling out, resulting in a momentum-independent amplitude. However, the interaction topology dictated by the $X$ and $\rho$ charge assignments under $U(1)_{\rm B-L}$ precludes this interference, resulting in cross-sections $\sigma \propto 1/k_{\mathsmaller{\rm CM}}^2$, or $\propto 1/T^2$ upon thermal averaging. 
We emphasize that $k_\mathsmaller{{\rm CM}}$ is the momentum in the center-of-momentum frame; with one of the particles being relativistic, $k_{\mathsmaller{\rm CM}} \sim T$ in a thermal bath. 
Consequently, the rate at which $X_2$ downscatters to $X_1$, exhibits \emph{super-critical} scaling: $\mathcal{R}_{2 \rho^{(*)} \to 1 \rho^{(*)}} = n_{\rho^{(*)}} \langle  \sigma \vMol \rangle_{2 \rho^{(*)} \to 1 \rho^{(*)}} \propto T$ decreases more slowly than the Hubble, $H \propto T^2/M_{\rm Pl}$, with $M_{\rm Pl}$ being the Planck mass, as the universe expands. The scatterings $X_2 \rho^{(*)} \rightarrow X_1 \rho^{(*)}$ thus become efficacious at sufficiently low temperatures, regardless of the coupling strength, as long as $\rho$ remains relativistic. This distinctive dynamics drives explosive asymmetry generation, which freezes out only when the $X_2$ population is sufficiently depleted.

The high rate of \emph{super-critical} scatterings may raise concerns about the validity of the perturbative expansion, suggesting that such interactions might necessitate resummation to properly account for thermal effects. Indeed, the interactions between the $X$ fermions and the $\rho$ bath induce temperature-dependent corrections to the $X$ mass matrix, potentially leading to large mixing between the zero-temperature mass eigenstates, particularly if they are closely spaced. This can affect the asymmetry generation and must be carefully addressed. In the regime of interest, the $X$ particles are highly non-relativistic, therefore their masses cannot be neglected, as often done in thermal field theory calculations. In \cref{App:ThermalEffects}, we show that thermal corrections generate off-diagonal terms $\propto T^2/\mX$ in the $X$ mass matrix. This matrix is diagonalized by a unitary transformation that evolves with temperature according to \cref{eq:mixing_angle}. The thermal mixing remains approximately negligible when
\begin{align}\label{eq:ThermCorrectCondition}
\dfrac{\Delta \mX}{\mX} \gg
\frac{T^2}{24\mX^2} |\y_{11} \y_{12}^* + \y_{22}^* \y_{12}|. 
\end{align}
This condition is satisfied for all $T \lesssim \mX$ rather comfortably, if $\Delta \mX/ \mX \gg |\y_{11} \y_{12}^* + \y_{22}^* \y_{12}|/24$. \\

\textbf{Lepton violation via critical scatterings~\textemdash  } 
In our model, the $X (\rho)$ violation occurs via $X_2 \rightarrow \bar{X}_1 \rho $ decays, and $X_i X_j \leftrightarrow \rho V_\mu$, $X_i V_\mu \leftrightarrow \bar{X}_j \rho $ scatterings. The decays are inherently efficient, provided that $m_\rho (T) < \Delta \mX$. Charge-2 annihilations are the least significant: their tree-level amplitude is suppressed for small $\aBL$, while for large $\aBL$ they become Sommerfeld suppressed at $\vrel \lesssim \aBL$ due to the repulsive $V_\mu-$mediated $XX$ potential. In contrast, charge-1 processes are enhanced at low temperatures: 
their cross-sections scale as $\sigma \propto 1/k_{\mathsmaller{\rm CM}}$,  or $\propto 1/T$ after thermal averaging. The difference compared to the CP-violating \emph{super-critical} scatterings discussed above is due to the momentum-suppressed fermion-fermion-vector vertex in place of one fermion-fermion-scalar vertex. 
The corresponding interaction rates, $\mathcal{R}_{i V_\mu \to \bar{j} \rho} = n_{V_\mu} \langle \sigma \vMol \rangle_{i V_\mu \to \bar{j} \rho} \propto T^2/\mX$, enhanced by the large density of a relativistic incoming species, scale with temperature in the same way as the Hubble rate. The $X_i V_\mu \leftrightarrow \bar{X}_j \rho $ interactions thus exhibits \emph{critical} scaling: provided that the relevant couplings are large enough, they reach and remain in equilibrium.
The efficient depletion of the $X_2$ species via these \emph{critical} scatterings, as well as the $X_2 \rightarrow \bar{X}_1 \rho$ decays, warrants that the $\Delta_\rho$ asymmetry eventually freezes out. 
\\

\textbf{Boltzmann equations \textemdash  }
We track the evolution of asymmetries by solving a system of coupled Boltzmann equations (BEqs) governing the abundances of $X_{1,2}, \bar{X}_{1,2}, \rho$ and $\rho^*$. We consider the yields $Y_l \equiv n_l/s$, where $n_l$ is the number density of species $l$, and $s$ is the entropy density of the universe. The equilibrium yields in the absence of chemical potential are denoted by $Y_l^{\rm eq}$. With this, the BEqs read
\begin{widetext}
\begin{subequations}
\label{eq:Boltzmann}
\label[pluralequation]{eqs:Boltzmann}
\begin{align}
    \label{eq:Boltzmann_YX}
    \frac{d Y_{i}}{dx} &= \dfrac{ds/dx}{3 H(x)} \bigg\{ 
    \mathcal{D}_{i}
    + [X_i V_\mu \leftrightarrow \bar{X}_j \rho]^j 
    + [X_i \rho^* \leftrightarrow \bar{X}_j V_\mu]^j
    + 2^{\delta_{ij}}[X_i X_j \leftrightarrow \rho V_\mu]^j 
    + \langle \sigma \vMol
\rangle_{i\bar{j}\to VV} 
      [Y_i Y_{\bar{j}} - Y_{i}^{\rm eq} Y_{\bar{j}}^{\rm eq}] 
    \\
    &
    + [X_i \bar{X}_j \leftrightarrow X_{i'\neq i} \bar{X}_{j'} ]^{ji'j'} 
    + [X_i \bar{X}_j \leftrightarrow \rho \rho^*]^{j}  
    + [X_i X_j \leftrightarrow X_{i^\prime} X_{j^\prime}]^{j i' j'}  
      \xi_{ij,i'j'}
    +[ X_i \rho^{(*)} \leftrightarrow X_{j\neq i} \rho^{(*)}]^j 
    \bigg\},
    \nn \\
    \label{eq:Boltzmann_Yrho}
    \frac{d Y_\rho}{dx} &= \dfrac{ds/dx}{3 H(x)} \bigg\{ 
    \mathcal{D}_{\bar{1}}  
    -\![X_i X_j \leftrightarrow \rho V_\mu]^{ij}  
    -\![ X_i V_\mu \leftrightarrow \bar{X}_j \rho]^{ij} 
    -\![X_i \bar{X}_j \leftrightarrow \rho \rho^*]^{ij} 
    +\!\langle \sigma \vMol \rangle_{\rho \rho^* \to {\rm rad}} 
      [Y_\rho Y_{\rho^*} - (Y_{\rho}^{\rm eq})^2]  
    \bigg\},
\end{align}
\end{subequations}
\end{widetext}
where $x \equiv m_1/T$, $\langle \cdot \rangle$ denotes the thermal average, including symmetry factors for the initial state, $[\cdot]^j$ indicates a summation over $j$, and 
\begin{align}
&\mathcal{D}_2 \equiv \dfrac{\langle \Gamma_{2 \to  \rho \bar{1}} \rangle}{s} 
\left(Y_2 - Y_2^{\rm eq} \ \dfrac{Y_{\rho} \, Y_{\bar{1}}}{Y_{\rho}^{\rm eq} Y_{\bar{1}}^{\rm eq}} \right), 
\quad {\cal D}_1 \equiv -{\cal D}_{\bar{2}},
\\
&[AB \!\leftrightarrow\! CD] 
\!\equiv\! 
\langle \sigma \vMol \rangle_{\!\mathsmaller{AB \to CD}} 
Y_\mathsmaller{A} Y_\mathsmaller{B} 
\!-\! 
\langle \sigma \vMol \rangle_{\mathsmaller{CD \to AB}} 
Y_\mathsmaller{C} Y_\mathsmaller{D},
\nonumber 
\end{align}
with $\vMol$ being the M{\o}ller velocity. Above, we have introduced the multiplicity factors $\xi_{iji'j'}$ with $\xi_{1112} = \xi_{1222} = \xi_{2212} = \xi_{2111}  = 1$, $\xi_{1122} = \xi_{2211} = 2$ and all other combinations vanishing. 
The last term in \cref{eq:Boltzmann_Yrho} includes $\rho\rho^{*}$ annihilations into $V_\mu V_\mu$ and other SM particles (e.g.,~due to the $\lambda_{\rho{\cal H}}$ coupling), and is assumed large enough to keep $\rho$ in chemical equilibrium with the plasma during asymmetry generation.  The BEqs for $Y_{\bar{i}}$ and $Y_{\rho^*}$ follow by conjugating \cref{eqs:Boltzmann}. The rates and cross-sections are provided in \cref{App:CrossSections}. The thermal averages are computed using standard methods; we refer to~\cite{Gondolo:1990dk, Davidson:2008bu, Belanger:2018ccd} for more details.
\\ 

Considering the conservation law, $\Delta_X = -2 \Delta_\rho$, and assuming rapid $\rho\rho^*$ annihilations that relate the $\rho$ and $\rho^*$  chemical potentials, $\mu_\rho = - \mu_{\rho^*}$, only four of the six of the above Boltzmann equations are independent. To gain insight into the dynamics of asymmetry generation, we introduce the following variables: 
$\beta \equiv Y_1-Y_2$, 
$\bar{\beta} \equiv Y_{\bar{1}}-Y_{\bar{2}}$, 
$\gamma \equiv Y_1+Y_2$, 
$\bar{\gamma} \equiv Y_{\bar{1}}+Y_{\bar{2}}$, 
with the $\rho$-number asymmetry being 
$\Delta_{\rho} = -\Delta_X/2 = (\bar{\gamma} - \gamma)/2$. The evolution of $\Delta_\rho$ is controlled by the following equation:
\begin{align}
\label{eq:delta_rhoEvolution}
\dfrac{d \Delta_\rho}{dx} = 
\dfrac{ds/dx}{3 H(x)} \bigg[ 
F_{\mathsmaller{\rm WO}} \cdot  \Delta_\rho + 
F_{\rm flavor} \cdot \dfrac{\beta - \bar{\beta}}{2} 
\bigg],
\end{align}
where $F_{\mathsmaller{\rm WO}}$, $F_{\rm flavor}$ depend on the interaction cross-sections and the yields, with their full expressions given in \cref{eq:Ffunctions}.
Noting that $F_{\mathsmaller{\rm WO}}>0$ implies that $F_{\mathsmaller{\rm WO}} \cdot \Delta_\rho$ can only washout the asymmetry. (We recall that $ds/dx <0$.) The term that sources the asymmetry is proportional to the flavor difference, $\beta - \bar{\beta}$, that encompasses the effect of CP violation. 
The evolution of $\beta - \bar{\beta}$ is governed by
\begin{align}
\dfrac{1}{2}
\dfrac{d (\beta - \bar{\beta})}{dx} \!
&\supset \! \dfrac{ds/dx}{3 H(x)} 
\Bigg\{\! \!- \!\langle \sigma \vMol \rangle_{2 \rho \rightarrow 1 \rho}^{\rm tree}    
\big(\varepsilon_{2 \rho \to 1 \rho} - \varepsilon_{\bar{2} \rho \to \bar{1} \rho}\big) \times \nonumber \\
&\times  Y_\rho^{\rm eq}\Bigg[ \!
\dfrac{\gamma + \bar{\gamma}}{2} \left(\!1+\frac{Y_2^{\rm eq}}{Y_1^{\rm eq}} \!\right) \!
-\dfrac{\beta + \bar{\beta}}{2} \left(\!1-\frac{Y_2^{\rm eq}}{Y_1^{\rm eq}}\! \right)  \! \!
\Bigg]  
\nonumber \\
&
+ {\cal O} (\Delta_\rho)
+ {\cal O} (\beta - \bar{\beta})
\Bigg\} .
\label{eq:FlavorAsymm_Evolution}
\end{align}
For simplicity, we have included above only the dominant \emph{super-critical} CP-violating scatterings, although all interactions are accounted for in the numerical computations. The various interactions contribute to the terms proportional to $\Delta_\rho$ and $\beta - \bar{\beta}$, but only CP-violating interactions appear in the combinations $\gamma + \bar{\gamma}$ and $\beta + \bar{\beta}$. Starting from symmetric initial conditions, $\Delta_\rho = \beta - \bar{\beta} = 0$, these terms proportional to  $\gamma + \bar{\gamma}$ and $\beta + \bar{\beta}$ generate a flavor asymmetry, which is subsequently converted into a $\rho$ asymmetry following \cref{eq:delta_rhoEvolution}.

At early times, the terms proportional to $\gamma + \bar{\gamma}$ dominate and seed the asymmetry. Considering that $(ds/dx)/H \propto - 1/x^2$ and $\langle \sigma \vMol \rangle_{2 \rho^{(*)} \to 1 \rho^{(*)}} \propto x^2$, the flavor asymmetry grows rapidly, feeding into \cref{eq:delta_rhoEvolution} to generate $\Delta_\rho$. 
This rapid growth is eventually curtailed by the depletion of the $X_2$ population (both equilibrium and actual) at $T < \Delta \mX$, which drives the factor inside the square brackets to zero. By contrast, for sub-critical interactions, the scaling of the cross-section suppresses the growth earlier, reducing the overall efficiency of asymmetry generation.
\\

\begin{table}[h!]
\begin{ruledtabular}
\begin{tabular}{cclcc}
$m_{1}$
& $\Delta \mX/\mX$ 
& $\{\alpha_{{{\rm B-L}}}, \alpha_{11}, \alpha_{22}, \alpha_{12}\}$ 
& $\Theta$ 
& $\w_{ij}$
\\
\hline\rule{0pt}{2.5ex}
$30$~PeV    
& $0.2$  
& $\{10^{-5},\, 10^{-1},\, 10,\, 10\}\times 10^{-3}$    
& $\pi/4$
& 0
\end{tabular}
\caption{Parameters used in~\cref{fig:AsymPlot}. We choose $\alpha_{11}, \alpha_{\rm B-L} \ll \alpha_{12}$ to prevent washout of the $X$ asymmetry. The remaining couplings, affecting the $\rho$ mass, are assumed to allow the $X_2 \to \bar{X}_1 \rho$ decays for $m_1/T \geqslant 1$ (cf.~\cref{eq:m_rho}).}
\label{tab:params}
\end{ruledtabular}
\end{table}

\begin{figure}[t!]
\centering
\includegraphics[width=0.97\linewidth]{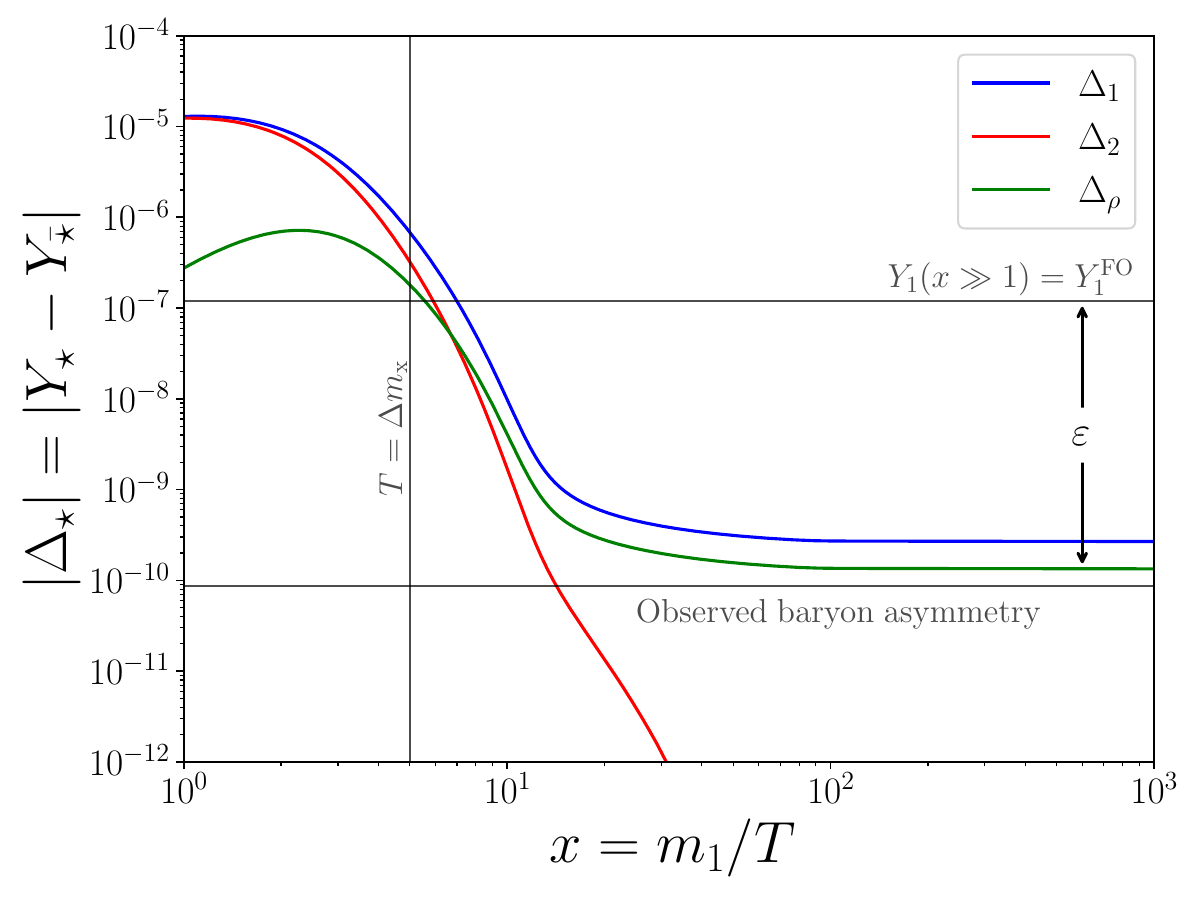}
\caption{\label{fig:AsymPlot}
Evolution of asymmetries for the model parameters in~\cref{tab:params}. $\Delta_1 + \Delta_2 = -2\Delta_\rho$ holds throughout. The asymmetry freezes-out at $T \lesssim \Delta \mX$ when $X_2,\bar{X}_2$ become underabundant due to downscatterings and/or decays into $X_1,\bar{X}_1$. The $\rho$ asymmetry is approximately $|\Delta_\rho|\sim |\varepsilon| Y_{1}^{\rm FO}$. }
\end{figure}

\textbf{Results \textemdash  }  While we do not perform a full parameter scan, we illustrate the dynamics of \emph{critical} and \emph{super-critical} processes in our model using the benchmark parameters in \cref{tab:params}, which yield an asymmetry near the observed value. For simplicity, we set $\w_{ij} = 0$, restricting the CP violation to the \emph{super-critical} scatterings and the long-range $X\bar{X}\leftrightarrow X\bar{X}$ processes.

The $X$-violating \emph{critical} scatterings $X_i V \leftrightarrow \bar{X}_j \rho$ and their conjugates contribute both to asymmetry generation and washout. The $X_2 V \leftrightarrow \bar{X}_1 \rho$ transitions are self-regulating: they become inefficient at $T \lesssim \Delta \mX$, in both directions, due to $X_2$ depletion and the energetic cost of upscattering. In contrast, $X_1 V_\mu \leftrightarrow \bar{X}_1 \rho$ scatterings remain a persistent source of washout. Their contribution to the evolution of the $X$ asymmetry is roughly
\begin{align}
\left(
\frac{d \ln \Delta_X}{d \ln x}
\right)_{1V \leftrightarrow \bar{1}\rho}
\approx
- \alpha_{11} \aBL
\frac{M_{\rm Pl}}{m_1} ,
\end{align}
continuously erasing any accumulated $X$ charge; this confirms their \textit{critical} behavior. To prevent washout, we choose small $\alpha_{11}$ and $\aBL$, noting that for these parameters, thermal corrections to $\mX$ are negligible, as per~\cref{eq:ThermCorrectCondition}.

\Cref{fig:AsymPlot} shows the evolution and freeze-out of the $\Delta_1$, $\Delta_2$, and $\Delta_\rho$ asymmetries. The large initial $X_1$ and $X_2$ yields, combined with the fact that equilibrium is never fully reached due to the cosmic expansion, allow sizable $|\Delta_1|$ and $|\Delta_2|$ to develop. However, they develop opposite signs, nearly canceling each other and resulting in a much smaller $|\Delta_\rho|$. As the effect of \emph{super-critical} CP-violating scatterings accumulates, a significant flavor asymmetry builds up, driving the growth of $|\Delta_\rho|$. The steep decline in all asymmetries at $T \lesssim \mX/3$ reflects the exponential suppression of the $X$ yields prior to freeze-out. At $T \lesssim \Delta\mX$, $X_2$ and $\bar{X}_2$ are efficiently converted into $X_1$ and $\bar{X}_1$ via decays and downscatterings, and the asymmetry freezes-out.

As previously mentioned, the yield of the non-relativistic particles around freeze-out establishes an upper bound on the final asymmetry. We observe numerically that 
\begin{align}
|\Delta_\rho| 
\simeq |\varepsilon| \, Y_{1}^{\rm FO}
\simeq
|\varepsilon| \, Y_{1}^{\rm eq}(x \sim 30),
\end{align}
which indicates that the asymmetry generation proceeds with nearly maximal efficiency, owing to the \emph{super-critical} processes. \\

\textbf{Discussion \textemdash  } 
We have identified a new class of CP-violating and/or particle-number-violating scattering processes in the early universe whose rates can exceed the Hubble expansion, providing an exceptionally efficient mechanism for generating the matter-antimatter asymmetry. This prodigious behavior arises from a combination of scalar interactions, which are not momentum suppressed, and a symmetry structure imposed by gauge invariance. Scalar fields, essential for CP violation, find a natural \emph{raison d'être} in this framework through their gauge charges, a structure that suggests a predestined role in cosmic asymmetry generation. 

Our findings reveal that \emph{critical} and \emph{super-critical} energy scalings of the cross-sections can significantly lower the scale at which asymmetry is produced. This has direct model-building and observational implications: the required couplings and CP phases may differ from standard expectations, and the energy regime relevant for baryogenesis may be brought closer to experimental reach. This mechanism opens a promising new direction for baryogenesis, leptogenesis, and asymmetric DM, warranting deeper theoretical and phenomenological investigation.

\begin{acknowledgments}
We are grateful to Sacha Davidson for pointing out the potential issue of thermal masses and for many insightful discussions. 
This work was supported by the European Union’s Horizon 2020 research and innovation programme under grant agreement No 101002846, ERC CoG CosmoChart.
\end{acknowledgments}

\bibliography{apssamp}

\clearpage
\appendix
\section*{Appendices}
\section{\label{app:Cosmo} Cosmological history}

\subsection{Phase transition and $\varrho$ decays} 

The $B-L$ symmetry is restored at high temperatures by thermal corrections to the potential (see e.g., \cite{Katz:2014bha}), which give $\rho$ the effective mass,
\begin{align}
m_\rho^2(T) \approx \left( \gBL^2  + \frac{\lambda_\rho }{12} + \frac{\lambda_{\rho \mathcal{H}}}{6} - \sum_{k=1}^3 \frac{|y_k^{\rm \chi}|^2}{4} \right) T^2 - \mu_\rho^2, 
\label{eq:m_rho}
\end{align} 
where we have included contributions from the gauge bosons, self-interactions, Goldstone modes, as well as the anomaly-canceling $\chi$ fermions in the Type A scenario (cf.~\cref{eq:rhoNR}). The $X$ fermions do not contribute to $m_\rho^2$, as they are non-relativistic at the relevant temperatures.  

At $T \lesssim T_{\cancel{\rm B-L}}$, where $T_{\cancel{\rm B-L}}$ can be estimated by setting $m_\rho^2 \to 0$, $\rho$ acquires a vacuum expectation value and breaks the local $B-L$ symmetry. 
Setting $\rho = (\vBL + \varrho)/\sqrt{2}$ in the unitary gauge, with
$\vBL \equiv 2\mu_\rho/ \lambda_\rho^{1/2}$, the $B-L$ phase transition endows the $\varrho$ and $V_\mu$ bosons with masses $m_\varrho = \sqrt{2} \mu_\rho$ and $m_V = 2 \gBL \vBL$. The LEP II experiment has set $m_V/\gBL = 2\vBL > 6 \, \rm{TeV}$~\cite{Carena:2004xs, Cacciapaglia:2006pk}. 

The massive $\varrho$ bosons can decay. 
The $\varrho \rightarrow V_\mu V_\mu$ mode is kinematically allowed if $\lambda_\rho > 32 \gBL^2$. Alternatively, the $\varrho$ bosons can decay into SM particles via off-shell $V_\mu$ and on- or off-shell Higgs bosons. The $\varrho$ decays dissipate the global charge that compensated the $X$ asymmetry during the $B-L$ symmetric phase, leaving the universe with a net $B-L$ charge stored in the particle sector.

\subsection{Cascade of the asymmetry}

The $X_1$ fermions decay into SM leptons with rate
$\Gamma_{X_1 \rightarrow L_l {\cal H}} \simeq |\lX_{1l}|^2 m_1/(16\pi)$.
The global charge carried by $X_1$ must cascade into SM leptons before the EW phase transition, such that it is reprocessed into a baryon number by sphalerons. However, after the $B-L$ breaking, the $X$ fermions acquire a small Majorana mass (see below). 
To prevent this from erasing the asymmetry while $X_1$ is decaying, we shall require that $X_1$ decays before the $B-L$ phase transition, 
$\Gamma_{X_1 \rightarrow L_i {\cal H} } \gtrsim H_{\cancel{\rm B-L}}$, where $H_{\cancel{\rm B-L}}$ denotes the Hubble rate at the $B-L$ breaking, or
\begin{align}
|\lX_{1l} | \gtrsim 10^{-8} 
\left(\dfrac{T_{\cancel{\rm B-L}}}{10~{\rm TeV}}\right) 
\left(\dfrac{100~{\rm PeV}}{m_1}\right)^{1/2} .
\label{eq:lambdaX}
\end{align}
The small values of $\lX_{1l}$ allowed by \cref{eq:lambdaX} warrant that any interactions introduced by $\lX_{1l}$ can be neglected during the asymmetry generation.  
\\

\begin{table}[t!]
\begin{ruledtabular}
\begin{tabular}{ccccc}
& $X_i$ & $\rho$ &$V_\mu$ &$\chi_k$ \\
\hline
$U(1)_{\rm B-L}$   
& $-1$ & $-2$ &$0$ 
& \makecell[c]{A: $\{-1, -1, -1\}$ \\ B: $\{+5, -4, -4\}$} 
\\
$\mathbb{Z}_2$
&+ &+ &+
& A: $-$~~~B: $+$ 
\\ \hline
mass {\phantom{$\int$}} & $\sim$ PeV & $\sim$ 10 TeV & $\sim$ TeV & 
\\
\end{tabular}
\caption{Charge assignments and indicative mass scales.}
\label{tab:Model}
\end{ruledtabular}
\end{table}

\subsection{Stable relics} 

In both Type A and B scenarios, summarized in \cref{tab:Model}, the anomaly-canceling fermions $\chi$ are stable relics, but with distinct cosmological roles. 
\\[1em]
\textit{Type A:} The $\chi$ fermions may couple to the $\rho$ boson via
\begin{align}
\mathcal{L}_{\rho,\chi} = - \sum_{k=1}^3  \frac{y^{\chi}_k}{2} \rho^\dagger \overline{\chi}_k^c \chi_{k}^{}  + h.c., \label{eq:rhoNR}
\end{align}
where we have diagonalized the $y^\chi$ matrix. The $\mathbb{Z}_2$ symmetry under which the $\chi$s are odd, precludes any other renormalizable couplings, including mixing with the $X$ fermions.
The operator \eqref{eq:rhoNR} generates $\chi$ masses after the 
$B-L$ breaking. Since the $\chi$s are massive and stable, the lightest among them could explain DM. If $\gBL$ and $y^\chi$ are large enough, the $\chi$ fermions thermalize in the early universe via annihilations into $V_\mu V_\mu$, $\rho\rho^*$, and SM particles. Once their interaction rate falls below the expansion rate, their comoving density freezes out. To ensure that their present abundance does not exceed $\Omega_{\rm DM}h^2 \approx 0.12 $ \cite{Planck:2018vyg}, $y^\chi$ and $\gBL$ must be sufficiently large. Conversely, for small $\gBL$ and $y^\chi$, the $\chi$ fermions remain out of equilibrium, but a significant abundance can still arise via pair creation from $V_\mu V_\mu$, $\rho\rho^*$, and SM particles (freeze-in). In this case, $\gBL$ and $y^\chi$ must be small enough to prevent overproduction. A detailed relic density calculation is beyond the scope of this work.
\\[1em] 
\textit{Type B:} The $B-L$ charges of the $\chi$ fermions forbid all interactions, including the operator \eqref{eq:rhoNR}, except their coupling to $V_\mu$. Consequently, the $\chi$s remain massless. They kinetically decouple once the $V_\mu$ bosons acquire mass, and evolve isentropically. The subsequent decoupling of the SM particles causes the $\chi$ temperature to redshift relative to the CMB, ensuring that their contribution to the relativistic energy of the universe comfortably satisfies the $\Delta N_{\rm eff} \lesssim 0.34 \, (2\sigma)$ bound \cite{Planck:2018vyg}.
\\

\subsection{Neutrino masses}
The model can also account for the active neutrino masses via an $X$-induced inverse seesaw mechanism~\cite{Mohapatra:1986aw,Mohapatra:1986bd}. After $B-L$ and EW symmetry breaking, the one-generation mass mixing matrix, in the $(\nu_{\rm L}, X_{\rm L}, X_{\rm R}^c)^T$ basis, is 
\begin{align}
\mathbb{M}_{F} =   
\begin{pmatrix} 
0 &0 & \lX \vEW / \sqrt{2}
\\
0 & y^{\rm L} \vBL / \sqrt{2} &\mX 
\\
{\lX}^* \vEW / \sqrt{2} 
&\mX 
& y^{\rm R *} \vBL / \sqrt{2} 
\end{pmatrix}.
\label{eq:Mmatrix_Neutrinos}
\end{align}
In the limit $\mX \gg \vBL, v_{\mathsmaller{\rm EW}}$, the diagonalization of \eqref{eq:Mmatrix_Neutrinos} yields a pair of Majorana fermions, consisting predominantly of $X_{\rm L}$ and $X_{\rm R}^c$, with masses around $\mX$ and a small mass splitting $\sim \vBL (y^{\rm L} + y^{\rm R *})/\sqrt{2}$.
The third eigenstate, consisting mainly of $\nu_{\rm L}$, has mass
\begin{align}
m_{\nu} \approx 
\frac{1}{2 \sqrt{2}} y^{\rm L}
|\lX|^2 
\frac{\vBL \, \vEW^2}{\mX^2}. 
\end{align}
This result holds both in Type A and B scenarios, in neither of which do the $\chi$ fermions mix with $\nu_L$ and $X_{\rm L, R}$.

\section{\label{App:CrossSections} Interactions and rates}

\subsection{Decays} 
The $X_2$ fermion, being the heaviest field, decays into $\bar{X}_1$ and $\rho$. For $\Delta \mX \ll \mX$, $\rho$ is emitted with energy $E_\rho \simeq \Delta \mX$, and the corresponding decay rate is
\begin{align}
\Gamma_{2 \rightarrow \rho \bar{1}} \simeq 
2\alpha_{12} \, \Delta \mX
\left[ 1- \left(\frac{m_\rho}{\Delta \mX} \right)^2\right]^{1/2},
\end{align}
where the temperature-dependent mass $m_\rho$ is found in \cref{eq:m_rho}.
The thermally-averaged decay rate that appears in the Boltzmann \cref{eqs:Boltzmann}, is given by 
\begin{align}
&\langle \Gamma_{2 \rightarrow \rho \bar{1}} \rangle 
\simeq \frac{K_1(x)}{K_2(x)} 
\left(1+ \dfrac{1}{e^{\Delta \mX/T}-1}\right)
\times 
\Gamma_{2 \rightarrow \rho \bar{1}}, 
\end{align}
for $x \gg 1$, where $K_n(x)$ are the modified Bessel functions of the second kind. The second factor in the parentheses above accounts for the Bose enhancement due to the final-state $\rho$ that takes away only a small amount of energy; it departs significantly from 1 for $\Delta \mX  \lesssim T$. On the other hand, decays are kinematically allowed at sufficiently low temperatures when $\Delta \mX > m_\rho$. 
\\

\subsection{Scatterings} 
The tree-level cross-sections, $\sigma^{\rm tree}$, for 2-to-2 processes --- excluding the interactions $X_i \bar{X}_j \leftrightarrow X_{i'} \bar{X}_{j'}$, which will be discussed in a companion paper~\cite{Flores:2025b} --- are listed in \cref{tab:CrossSections} and grouped according to the total $B-L$ charge of the interacting states. We consider $s$-wave contributions only. For processes involving non-relativistic $XX$ or $X\bar{X}$ pairs, which interact via long-range potentials, we also include the relevant Sommerfeld factors. In more detail:

$\bm{XX \rightarrow XX}$: 
$XX$ pairs interact via soft \(V_\mu\) boson exchanges, which do not mix flavor states. However, hard $s$-channel $\rho$ exchange results in flavor-changing scatterings. 
The repulsive $V_\mu-$mediated potential introduces Sommerfeld suppression factors both due to the incoming $(\mathcal{S})$ and outgoing $(\mathcal{S}')$ states, 
$S_0(-\zeta_{\mathsmaller{\rm B-L}})
S_0(-\zeta_{\mathsmaller{\rm B-L}}')$,  
with 
\begin{align} 
\label{eq:Sommerfeld}
S_0(\zeta) = \frac{2 \pi \,\zeta}{1- e^{-2 \pi\,  \zeta}},
\end{align}
and $\zeta_{\mathsmaller{\rm B-L}}^{(\prime)} \equiv \aBL / \vrel^{(\prime)}$, 
where the superscript $^\prime$ refers to the outgoing states. 
The relative velocities of the incoming and outgoing states are related via
    \begin{align}
s \simeq 
\left(m_{\cal S}^{} +\dfrac{{\bf k}_{\cal S}^2}{2\mu_{\cal S}^{}} \right)^2 
\simeq 
\left(m_{{\cal S}'}^{} +\dfrac{{\bf k}_{{\cal S}'}^2}{2\mu_{{\cal S}'}^{}} \right)^2,
\label{eq:s_vs_kk'}
\end{align} where $|{\bf k}_{{\cal S}^{(\prime)}}| = \mu_{\cal S^{(\prime)}}\vrel^{(\prime)}$ is the momentum of the incoming (outgoing) states in the center-of-momentum (CM) frame, and $\mu_{\mathcal{S}^{(\prime)}}$ denotes the reduced mass. At leading order in $\Delta \mX/\mX$, the relative velocities are approximately equal, i.e., $v_{\rm rel}^\prime \approx v_{\rm rel}$.
\\

\textbf{$\bm{X\rho^{(*)} \rightarrow X\rho^{(*)}}$ scatterings:} 
Let us briefly comment on the divergences that arise in diagrams mediated by the $V_\mu$ boson. These include both tree-level elastic scatterings, e.g., $X_i \rho^{(*)} \to X_i \rho^{(*)}$, and loop-level processes, such as diagrams (iii) in \cref{fig:CPdiag3,fig:CPdiag1}. We illustrate this issue using tree-level elastic scatterings as an example.
The $X_i \rho^{(*)} \to X_i \rho^{(*)}$ interactions are mediated in the $u-$channel by the $X_j$ exchange, and in the $t-$channel by $V_\mu$, resulting in the total amplitude $\mathcal{A}_{i \rho^{(*)} \to i \rho^{(*)}} = \mathcal{A}_u + \mathcal{A}_t$. The $\mathcal{A}_u$ part provides a finite contribution to the cross-section. However, the t-channel amplitude behaves as $\mathcal{A}_t \propto 1/(\cos{\theta}-1)$, and becomes singular in the limit $\theta \rightarrow 0$, where $\theta$ denotes the scattering angle. Consequently, both the interference term $\mathcal{A}_u \mathcal{A}_t^* + \mathcal{A}_u^* \mathcal{A}_t $, and the squared $t-$channel contribution $|\mathcal{A}_t|^2$ give rise to divergent angular integrals. To render the cross-section finite, we regularize the integral over $\theta$ by introducing a cutoff angle $\theta_{\min}$. The relevant integrals become
\begin{subequations}
\begin{align}
    \mathcal{I}_1 &\equiv\int_0^\pi \frac{\sin{\theta} \, d \theta}{\cos{\theta}-1} \rightarrow \int_{\theta_{\mathsmaller{min}}}^\pi \frac{\sin{\theta} \, d \theta}{\cos{\theta}-1} = 2 \ln{ \left(\sin{\frac{\theta_{\mathsmaller{ min }}}{2}} \right)}, \label{eq:I1int}\\
    \mathcal{I}_2 &\equiv \int_0^\pi \frac{\sin{\theta} \, d \theta}{(\cos{\theta}-1)^2} \rightarrow \int_{\theta_{\mathsmaller{min}}}^\pi \frac{\sin{\theta} \, d \theta}{(\cos{\theta}-1)^2} \nonumber \\
    &= \frac{1}{1-\cos{\theta_{\mathsmaller{\min}}}} - \frac{1}{2}.
\end{align}
\end{subequations}
Expanding around a small cutoff angle and retaining only the finite part, we find
\begin{align} \label{eq:reg2}
    &\mathcal{I}_1 \approx  - 2 \ln{2}, &\mathcal{I}_2 \approx  - \frac{1}{3}.
\end{align}
We encounter divergences of type $\mathcal{I}_1$ when calculating the CP-violating coefficients $\varepsilon$, which arise from the interference between the tree-level diagram and the one-loop diagram involving a $t-$channel exchange of $V_\mu$ (diagrams $(iii)$ in \cref{fig:CPdiag3,fig:CPdiag1}).

\section{\label{App:ThermalEffects} Thermal effects}
At finite temperature, the resummed fermion propagator can be written as \cite{Weldon:1982bn,Levinson:1985ub,Bellac:2011kqa}
\begin{align}
    S_F(p) = \frac{\im }{\slashed{p} - \Sigma(p) +\im \epsilon },
\end{align}
with $p$ being the external momentum, and $\Sigma(p)$ denoting the fermionic self-energy
\begin{align}
    \Sigma(p) = -a  \slashed{p} - b \slashed{u} + \mX,
\end{align}
where $a$ and $b$ are Lorentz invariant structure functions, which, in general, might depend on $p_0$ and $|\mathbf{p}|$, and $u$ is the four-velocity of the heat bath, normalized as $u \cdot u = 1$. In the rest frame of the heat bath, the structure functions are given by  
\begin{subequations}
\begin{align}
    &a(p_0, |\mathbf{p}|) = \frac{1}{4 |\mathbf{p}|^2} \left[ {\rm Tr}(\Sigma \slashed{p}) - p_0 {\rm Tr}(\Sigma \slashed{u}) \right], \\
    &b(p_0, |\mathbf{p}|) = \frac{1}{4 |\mathbf{p}|^2} \left[ p^2 {\rm Tr}(\Sigma \slashed{u}) - p_0 {\rm Tr}(\Sigma \slashed{p}) \right].
\end{align}
\end{subequations}
The propagator $S_F(p)$ has a pole at 
\begin{align}
    (1+a)^2 (p_0^2 - |\mathbf{p}|^2) + 2b(1 + a) p_0 + b^2 - \mX^2 = 0,
\end{align}
so that the dispersion relation can be written as
\begin{align}
    p_0 = \frac{-b \pm \left(\mX^2 + (1+a )^2 |\mathbf{p}|^2 \right)^{1/2}}{1+a},
\end{align}
and the mass shift as  \cite{Levinson:1985ub}
\begin{align}
    \delta m = \lim_{|\mathbf{p}| \rightarrow 0} p_0 - m. \label{eq:mass_shift}
\end{align}
\begin{figure}[h!]
\centering
\begin{tikzpicture}[line width=1.1pt, scale=1]
\begin{scope}
\node at (-3.1,.4) {$X_i$};
\node at (.5,.4) {$X_k$};
\node at (-2.3,.6) {\textcolor{gray}{$p$}};
\node at (-.25,.6) {\textcolor{gray}{$p$}};
\node at (-1,1.2) {\textcolor{gray}{$k$}};
\node at (-1.3,.6) {\textcolor{gray}{$p-k$}};
\node at (-1.3,.1) {$\bar{X}_j$};
\node at (-1.9,.85) {$\rho$};
\draw[scalar] ( -1.9,.4) arc (180:0:.6);
\draw[fermion]  ( -2.8,.4)--( -1.8,.4);
\draw[fermion]  ( -.8,.4)--( -1.8,.4);
\draw[fermion]  ( -.8,.4)--(.2,.4);
\end{scope}
\end{tikzpicture}
\caption{One-loop contribution of the self-energy of the $X$ field.} 
\label{fig:SEdiag}
\end{figure}
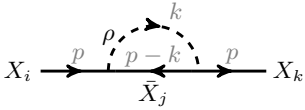
\subsection{Self-energy of the $X$ field}
The one-loop contribution to the self-energy of the $X$ field, shown in \cref{fig:SEdiag}, is given by 
\begin{align}
    \Sigma(p) &=  \int \frac{d^4 k}{(2 \pi)^4} \frac{\im }{k^2}   
    \im(y_{jk}^{\rm L *}P_{\rm R} + y_{jk}^{\rm R *}P_{\rm L})  \nonumber \\
    &\times \frac{\im(\slashed{p}- \slashed{k} + m_j)}{(p-k)^2 - m_j^2} \im (y_{ij}^{\rm L} P_{\rm L} + y_{ij}^{\rm R} P_{\rm R}).
\end{align}
In the imaginary time formalism, $\Sigma(p)$ can be expressed as
\begin{align}\label{eq:SE_IT}
    \Sigma(p) &\simeq T \displaystyle\SumInt_{k_0} \frac{1}{k^2} \frac{1}{(p-k)^2 - m_j^2} \nonumber \\
    &\times (y_{jk}^{\rm L *}P_{\rm R} + y_{jk}^{\rm R *}P_{\rm L}) (\slashed{p}-\slashed{k} + m_j) (y_{ij}^{\rm L} P_{\rm L} + y_{ij}^{\rm R} P_{\rm R}),
\end{align}
where $k_0 = 2 \pi i n T $, and $n \in \mathbb{Z}$. The relevant contractions read
\begin{subequations}
\begin{align}
    {\rm Tr}( \Sigma \slashed{p}) = 2 (y_{ij}^{\rm L}y_{jk}^{\rm L *} + y_{ij}^{\rm R}y_{jk}^{\rm R *} ) \times T \; \displaystyle \SumInt_{k_0} \frac{p^2 - p \cdot k}{k^2[(p-k)^2 - m_j^2]}, \\
    {\rm Tr}( \Sigma \slashed{u}) = 2 (y_{ij}^{\rm L}y_{jk}^{\rm L *} + y_{ij}^{\rm R}y_{jk}^{\rm R *} ) \times T \; \displaystyle \SumInt_{k_0}  \frac{p_0 - k_0}{k^2[(p-k)^2 - m_j^2]},
\end{align}
\end{subequations}
so that the structure functions become
\begin{subequations}
\begin{align}
    a(p_0, |\mathbf{p}|) \simeq \frac{(y_{ij}^{\rm L}y_{jk}^{\rm L *} + y_{ij}^{\rm R}y_{jk}^{\rm R *} ) }{2 |\mathbf{p}|^2} \times T  \displaystyle \SumInt_{k_0} \frac{\mathbf{p}\cdot \mathbf{k} - \mathbf{p}^2}{k^2[(p-k)^2 - m_j^2]},\\
    b(p_0, |\mathbf{p}|) \simeq \frac{(y_{ij}^{\rm L}y_{jk}^{\rm L *} + y_{ij}^{\rm R}y_{jk}^{\rm R *} ) }{2 |\mathbf{p}|^2} \times T  \displaystyle \SumInt_{k_0} \frac{|\mathbf{p}|^2 k_0 - p_0 \, \mathbf{p} \cdot \mathbf{k} }{k^2[(p-k)^2 - m_j^2]}.  \label{eq:B}
\end{align}
\end{subequations}
To evaluate the expressions above, we make use of the following Matsubara sums (see e.g., \cite{Mustafa:2022got}):
\begin{subequations}
\begin{align}
    \mathcal{M}_1 &\equiv T\sum_{k_0 = 2 \pi i n T} \frac{1}{k^2 [(p-k)^2 - m_j^2]}  \nonumber \\
    &=\sum_{s_1,s_2 = \pm} - \frac{s_1 s_2 }{4 \omega_k \omega_{p-k}} \frac{1+ f_{B} (s_1 \omega_k) -f_{F}(s_2 \omega_{p-k})}{p_0 - s_1 \omega_{k} - s_2 \omega_{p-k}}  \nonumber \\
    &=\frac{1}{4 \omega_k \omega_{p-k}} \bigg\{ [1 + f_{B}(\omega_k) - f_{F}(\omega_{p -k})] \times \nonumber \\
    &\times \left( \frac{1}{p_0 + \omega_k + \omega_{p-k}} - \frac{1}{p_0 - \omega_k - \omega_{p-k}}\right) \nonumber \\
    &+[f_{B}(\omega_k) + f_{F}(\omega_{p-k}) ] \times \nonumber \\
    &\times \left( \frac{1}{p_0 - \omega_k + \omega_{p-k}} - \frac{1}{p_0 + \omega_k - \omega_{p-k}}\right) \bigg\},
\end{align}
and 
\begin{align}
    \mathcal{M}_2 &\equiv T\sum_{k_0 = 2 \pi i n T} \frac{k_0}{k^2 [(p-k)^2 - m_j^2]}  \nonumber \\
    &=\sum_{s_1, s_2 = \pm} - \frac{s_2 }{4 \omega_{p-k}} \frac{1+ f_{B} (s_1 \omega_k) -f_{F}(s_2 \omega_{p-k})}{p_0 - s_1 \omega_{k} - s_2 \omega_{p-k}}  \nonumber \\
    &=\frac{-1}{4 \omega_{p-k}} \bigg\{ [1 + f_{B}(\omega_k) - f_{F}(\omega_{p -k})] \times \nonumber \\
    &\times \left( \frac{1}{p_0 + \omega_k + \omega_{p-k}} + \frac{1}{p_0 - \omega_k - \omega_{p-k}}\right) \nonumber \\
    &- [f_{B}(\omega_k) + f_{F}(\omega_{p-k}) ] \times \nonumber \\
    &\times \left( \frac{1}{p_0 - \omega_k + \omega_{p-k}} + \frac{1}{p_0 + \omega_k - \omega_{p-k}}\right) \bigg\},
\end{align}
\end{subequations}
where 
\begin{align}
   &\omega_k \equiv |\mathbf{k}|, &&\omega_{p-k} \equiv \sqrt{(\mathbf{p} - \mathbf{k})^2 + m_j^2},
\end{align}
and 
\begin{align}
    &f_{B}(\omega) = \frac{1}{e^{\omega/T}-1}, &&f_{F}(\omega) = \frac{1}{e^{\omega/T}+1}.
\end{align}
\subsection{Non-relativistic approximation}
We seek to find the $X$ mass matrix at a period relevant for the asymmetry generation. To that end, we compute the structure functions in the non-relativistic limit
\begin{align}
    p_0 = \sqrt{m_i^2 +|\mathbf{p}|^2} \gg |\mathbf{p}| \sim \sqrt{m_i T} \gg k_0, |\mathbf{k}| \sim T. 
\end{align}
Neglecting the mass difference, i.e., $m_i \approx m_j \equiv \mX$, we can make the following approximations:
\begin{subequations}
\begin{align}
    &\omega_{p-k} \approx p_0 - \frac{\mathbf{p} \cdot \mathbf{k}}{p_0}, \\
    &p_0 \pm \omega_k + \omega_{p-k} \approx 2 p_0 \pm |\mathbf{k}| - \frac{\mathbf{p \cdot k}}{p_0}, \\
    &p_0 \pm \omega_k - \omega_{p-k} \approx  \pm |\mathbf{\mathbf{k}}| 
    +
    \frac{\mathbf{p} \cdot \mathbf{k}}{p_0},  \\
    &f_{F}(\omega_{p-k}) \approx f_{F}(p_0) - \frac{f'(p_0)}{p_0} \mathbf{p}\cdot \mathbf{k} \overset{\mX \gg T}{\rightarrow} 0,
\end{align}
\end{subequations}
under which the Matsubara sums simplify as 
\begin{subequations}
\begin{align}
    \mathcal{M}_1 &\overset{\mX \gg |\mathbf{p}| \gg k }{\approx} \frac{1+ f_B( |\mathbf{k}|)}{4 |\mathbf{k}|^2 \mX} \bigg( 1 + \frac{\mathbf{p} \cdot \mathbf{\hat{k}}}{\mX}  \bigg) \nonumber \\
    &+\frac{f_B( |\mathbf{k}|)}{4 |\mathbf{k}|^2 \mX} \bigg(- 1 + \frac{\mathbf{p} \cdot \mathbf{\hat{k}}}{\mX}  \bigg), \\
  \mathcal{M}_2  &\overset{\mX \gg |\mathbf{p}| \gg k }{\approx} \frac{1+ f_B( |\mathbf{k}|)}{4 |\mathbf{k}| \mX} \bigg( 1 +  \frac{\mathbf{p} \cdot \mathbf{\hat{k}}}{\mX} - \frac{|\mathbf{k}|}{\mX}
    \bigg) \nonumber \\
    &+\frac{f_B( |\mathbf{k}|)}{4 |\mathbf{k}| \mX} \bigg(1 - \frac{\mathbf{p} \cdot \mathbf{\hat{k}}}{\mX} + \frac{|\mathbf{k}|}{\mX} \bigg).
\end{align}
\end{subequations}
Extracting the finite part of the structure functions, i.e., including the Bose-Einstein distribution, we get
\begin{align}
    a(p_0, |\mathbf{p}|) &\approx \frac{(y_{ij}^{\rm L}y_{jk}^{\rm L *} + y_{ij}^{\rm R}y_{jk}^{\rm R *} ) }{4 |\mathbf{p}|^2 \mX^2} \frac{1}{(2 \pi)^2} \int_{-1}^1 d \cos{\theta} \nonumber \\
    &\times \int_0^\infty d |\mathbf{k}| \bigg( |\mathbf{p}||\mathbf{k}| \cos{\theta} - |\mathbf{p}|^2\bigg) |\mathbf{p}| \cos{\theta} f_B(|\mathbf{k}|)  \nonumber \\
    &= \frac{(y_{ij}^{\rm L}y_{jk}^{\rm L *} + y_{ij}^{\rm R}y_{jk}^{\rm R *} ) }{144} \frac{T^2}{\mX^2},
    \end{align}
and 
\begin{align}
    b(p_0, |\mathbf{p}|) &\approx \frac{(y_{ij}^{\rm L}y_{jk}^{\rm L *} + y_{ij}^{\rm R}y_{jk}^{\rm R *} ) }{4 \mX} \frac{1}{(2 \pi)^2} \int_{-1}^1 d \cos{\theta} ( 1 - \cos{\theta}^2 ) \nonumber \\
    &\times \int_0^\infty d |\mathbf{k}| |\mathbf{k}|  f_B(|\mathbf{k}|) \nonumber \\
    &= \frac{(y_{ij}^{\rm L}y_{jk}^{\rm L *} + y_{ij}^{\rm R}y_{jk}^{\rm R *} ) }{72} \frac{T^2}{\mX}. 
    \end{align}
Then, from Eq.\eqref{eq:mass_shift} we find the mass shift
\begin{align}
    &\delta m_{ij} \approx 
    - \frac{Y_{ik}}{48} \frac{T^2}{\mX}, &Y_{ik} \equiv \sum_{j={\{1,2\}}} y_{ij}^{\rm L}y_{jk}^{\rm L *} + y_{ij}^{\rm R}y_{jk}^{\rm R *},
\end{align}
and by virtue of which the $X$ mass matrix becomes
\begin{align}
    \mathbb{M}_X \approx \begin{pmatrix}
m_1 - \frac{T^2}{\mX} \frac{Y_{11}}{24} & - \frac{T^2}{\mX} \frac{Y_{12}}{24} \\
- \frac{T^2}{\mX} \frac{Y_{12}^*}{24} & m_2 - \frac{T^2}{\mX} \frac{Y_{22}}{24}
\end{pmatrix},
\end{align}
The $\mathbb{M}_X$ matrix can be diagonalized by a unitary matrix, characterized by a mixing angle
\begin{align}
    \theta
    &= \frac{1}{2} \arctan{ \left( \frac{ x^{-2} |Y_{12}|}{24 \, \frac{\Delta \mX}{\mX} + x^{-2} \left(Y_{11} - Y_{22} \right)}\right)}. \label{eq:mixing_angle}
\end{align}
For the choice $\w_{ij} = 0$ and $|\y_{11}|^2 = |\y_{22}|^2$, the mixing angle becomes small when
\begin{align}
    24 \frac{\Delta \mX}{\mX} \gg \frac{1}{x^2} |Y_{12}|.
\end{align}
Since $x$ decreases with time, this condition imposes the strongest constraint on the coupling at $x \sim 1$, i.e., 
\begin{align}
    24 \frac{\Delta \mX }{\mX} \gg |\y_{11} \y_{12}^* + \y_{22}^* \y_{12}|. 
\end{align}

\clearpage
\section{\label{App:BoltzmannEquations} Boltzmann equations for the asymmetries}
The four independent Boltzmann equations can be reformulated in terms of the following dimensionless variables:
\begin{align}
&\gamma = Y_1 + Y_2, &&\bar{\gamma} = Y_{\bar 1} + Y_{\bar 2}, \\
    &\beta = Y_1 - Y_2, &&\bar{\beta} = Y_{\bar 1} - Y_{\bar 2}.
\end{align}
By inverting these relations and imposing the constraint $\mu_\rho = -\mu_{\rho^*}$ on the $\rho$ chemical potential, one obtains 
\begin{align}
    &Y_i = \frac{1}{2} \bigg(\gamma -(-1)^i \beta\bigg), &&Y_{\bar i} = \frac{1}{2}\bigg(\bar{\gamma} - (-1)^i \bar{\beta}\bigg),\\
      &Y_\rho \simeq   Y_\rho^{\rm eq} + \frac{\Delta_\rho}{2},
     &&Y_{\rho^*} \simeq  Y_\rho^{\rm eq} - \frac{\Delta_\rho}{2},
\end{align}
and 
\begin{align}
 \Delta_\rho &\equiv \frac{\bar{\gamma} - \gamma}{2}, \\
     \Delta_i &\equiv - \Delta_\rho - (-1)^i \frac{\beta - \bar{\beta}}{2}, && i \in \{1,2\}.
\end{align}
The Boltzmann equation for the total asymmetry then reads
\begin{align}
  \frac{d \Delta_\rho }{dx} &= \frac{ds/dx}{3 H(x)} \bigg\{ \mathcal{D}_{\bar 1} - \mathcal{D}_{1} \nonumber \\
   &+
   [\bar{X}_i \bar{X}_j \leftrightarrow \rho^* V_\mu]^{ij} - [X_i X_j \leftrightarrow \rho V_\mu]^{ij}  \nonumber \\
   &+ [\bar{X}_i V_\mu \leftrightarrow X_j \rho^*]^{ij}  -[X_i V_\mu \leftrightarrow \bar{X}_j \rho]^{ij} \bigg\}.
\end{align}
By simple rearrangements, the right-hand side of the above equation separates into a part proportional to
$\Delta_\rho$ and another proportional to $\beta - \bar{\beta}$, i.e., 
\begin{align}\label{eq:EqDeltarho}
    \frac{d \Delta_\rho}{dx } = \frac{ds/dx}{3 H(x)} \bigg[ F_{\mathsmaller{\rm WO}} \cdot  \Delta_\rho + F_{\rm flavor}\cdot \frac{\beta - \bar{\beta}}{2} \bigg],
\end{align}
where
\begin{widetext}
\begin{subequations}\label{eq:Ffunctions}
    \begin{align}
         F_{\mathsmaller{\rm WO}}  &\equiv \frac{\langle \Gamma_{2 \to \rho \bar{1}} \rangle}{s} \bigg[ 1+ \frac{Y_2^{\rm eq}}{Y_1^{\rm eq}} + \frac{Y_2^{\rm eq}}{Y_\rho^{\rm eq} Y_1^{\rm eq}} \frac{Y_1+ Y_{\bar{1}}}{2} \bigg]+ \sum_{i,j} \langle \sigma \vMol \rangle_{ij \to \rho V_\mu} \bigg[ Y_i + Y_{\bar{j}} +  Y_{V_\mu}^{\rm eq} \frac{Y_i^{\rm eq} Y_{j}^{\rm eq}}{Y_\rho^{\rm eq} Y_{V_\mu}^{\rm eq}} \bigg] \nonumber \\
        &+ Y_{V_\mu}^{\rm eq} \sum_{ij} \langle \sigma \vMol \rangle_{iV_\mu \rightarrow \bar{j} \rho} \bigg[ 1+ \frac{ Y_i^{\rm eq}}{Y_j^{\rm eq} Y_\rho^{\rm eq}} (Y_j+ Y_\rho) \bigg], \\
        F_{\rm flavor} &\equiv \frac{\langle \Gamma_{2 \to \bar{1} \rho} \rangle}{s} \bigg[ 1 - \frac{Y_2^{\rm eq}}{Y_1^{\rm eq}}\bigg]+ \sum_{i,j} \langle \sigma \vMol \rangle_{ij \to \rho V_\mu} \left[Y_i (-1)^j + Y_{\bar{j}} (-1)^i \right] + Y_{V_\mu}^{\rm eq} \sum_{i,j} \langle \sigma \vMol \rangle_{iV_\mu \rightarrow \bar{j} \rho} \bigg[ (-1)^i + (-1)^j \frac{Y_\rho Y_i^{\rm eq}}{Y_j^{\rm eq} Y_\rho^{\rm eq}} \bigg].
    \end{align}
\end{subequations}
\end{widetext}
Due to the Sommerfeld or gauge coupling suppression, the $X_i X_j \rightarrow \rho V_\mu$ interactions contribute only marginally to the right-hand side of \cref{eq:EqDeltarho}. Consequently, the $\rho-$number asymmetry is primarily driven by the decays $X_2 \to \bar{X}_1 \rho $ and the \emph{critical} scatterings $X_i V_\mu \leftrightarrow \bar{X}_j\rho$. However, taking into account their temperature dependence --- $\langle \Gamma_{2 \to \rho \bar{1}} \rangle /s \sim x^{3}$ for decays and $\langle \sigma \vMol \rangle_{i V_\mu \to \bar{j} \rho} \sim x$ for scatterings --- it follows that decays eventually dominate the asymmetry generation at late times. \\
Owing to the fact that $F_{\mathsmaller{\rm WO}}$ remains non-negative throughout the evolution, the term $F_{\mathsmaller{\rm WO}} \cdot \Delta_\rho$ always tends to erase the asymmetry. Therefore, the generation of asymmetry requires a non-zero contribution from $\beta - \bar{\beta}$, which evolves according to the equation  \begin{align}\label{eq:Eqbeta}
        \frac{d (\beta- \bar{\beta})}{dx} &\simeq \frac{ds/dx}{3 H(x)} \Bigg( \mathcal{S}_1 
        + \mathcal{S}_2 + \mathcal{S}_3 + \text{higher-order terms} \bigg),
    \end{align}
where the source terms are defined in \cref{eq:source terms}. To reduce complexity, in the expression above, we have omitted contributions from long-range annihilations $[X_i \bar{X}_j \leftrightarrow \rho \rho^{(*)}]$ and flavour-changing scatterings $[X_i \bar{X}_j \leftrightarrow X_{i'} \bar{X}_{j'}]$, as well as charge-2 Sommerfeld-suppressed interactions $[X_i X_j \leftrightarrow X_{i'} X_{j'}]$. 
These processes involve at least one pair of non-relativistic particles and typically yield subleading contributions. \\ 
In contrast, the dominant source of CP asymmetry arises from \emph{super-critical} scatterings ($\mathcal{S}_3$), which generate the leading-order term $\sim -\frac{ds/dx}{3H(x)} \langle \sigma \vMol \rangle^{\rm tree}_{2 \rho \to 1 \rho} (\epsilon_{2 \rho \to 1 \rho} - \epsilon_{\bar{2} \rho \to \bar{1} \rho}) (\gamma + \bar{\gamma}) Y_\rho^{\rm eq}$ on the right-hand side of \cref{eq:Eqbeta}. The CP-preserving terms ($\mathcal{S}_1, \mathcal{S}_2$) vanish when both the $\rho-$number asymmetry and the flavor difference are zero. In contrast, the CP-violating interactions remain generally non-zero even when $\Delta_\rho =0$ and $\beta - \bar{\beta} =0$.  \\
\begin{widetext}
\begin{subequations} \label{eq:source terms}
\begin{align}
    \mathcal{S}_1 &\equiv [ X_1 V_\mu \leftrightarrow \bar{X}_j \rho]^j - [X_2 V_\mu \leftrightarrow \bar{X}_j \rho]^j - [\bar{X}_1 V_\mu \leftrightarrow X_j \rho^*]^j +  [\bar{X}_2 V_\mu \leftrightarrow X_j \rho^*]^j \nonumber \\
    &+ [X_1 \rho^* \leftrightarrow \bar{X}_j V_\mu]^j -  [X_2 \rho^* \leftrightarrow \bar{X}_j V_\mu]^j -  [\bar{X}_1 \rho \leftrightarrow X_j V_\mu]^j + [\bar{X}_2 \rho \leftrightarrow X_j V_\mu]^j \nonumber \\
    &= - 2 Y_{V_\mu}^{\rm eq} \bigg\{ \Delta_\rho \bigg[ \langle \sigma \vMol \rangle_{1 V_\mu \to \bar{1} \rho}\bigg( 2 + \frac{Y_1 + Y_{\bar 1}}{2 Y_\rho^{\rm eq}} \bigg) -\langle \sigma \vMol \rangle_{2 V_\mu \to \bar{2} \rho}\bigg( 2 + \frac{Y_2 + Y_{\bar 2}}{2 Y_\rho^{\rm eq}} \bigg) \bigg] \nonumber \\
    &- \frac{\beta - \bar{\beta}}{2}  \bigg( \langle \sigma \vMol \rangle_{1 V_\mu \to \bar{1} \rho} + \langle \sigma \vMol \rangle_{2 V_\mu \to \bar{2} \rho} \bigg) \bigg\}, \\ \nonumber \\
    \mathcal{S}_2 &\equiv  2^{
    1j}[X_1 X_j \leftrightarrow \rho V_\mu]^j -2^{
    2j} [X_2 X_j \leftrightarrow \rho V_\mu]^j - 2^{
    1j} [\bar{X}_1 \bar{X}_j \leftrightarrow \rho^* V_\mu]^j + 2^{
    2j} [\bar{X}_2 \bar{X}_j \leftrightarrow \rho^* V_\mu]^j \nonumber \\
    & =  - 2 \Delta_\rho \bigg[ \langle \sigma \vMol \rangle_{11 \to \rho V_\mu} \bigg( Y_1 + Y_{\bar{1}}  + \frac{Y_1^{\rm eq} Y_1^{\rm eq}}{Y_\rho^{\rm eq} } \bigg) -  \langle \sigma \vMol \rangle_{22 \to \rho V_\mu} \bigg( Y_2 + Y_{\bar{2}}  + \frac{Y_2^{\rm eq} Y_2^{\rm eq}}{Y_\rho^{\rm eq} } \bigg) \bigg] \nonumber \\
    &+(\beta-\bar{\beta}) \bigg[ \langle \sigma \vMol \rangle_{11 \to \rho V_\mu}  (Y_1 + Y_{\bar{1}}) + \langle \sigma \vMol \rangle_{22 \to \rho V_\mu}  (Y_2 + Y_{\bar{2}}) \bigg], \\ \nonumber \\
     \mathcal{S}_3 &\equiv [X_1 \rho^{(*)} \leftrightarrow X_2 \rho^{(*)}] - [X_2 \rho^{(*)} \leftrightarrow X_1 \rho^{(*)}] - [\bar{X}_1 \rho^{(*)} \leftrightarrow \bar{X}_2 \rho^{(*)}] + [\bar{X}_2 \rho^{(*)} \leftrightarrow \bar{X}_1 \rho^{(*)}] \nonumber \\
    &= 2\langle \sigma \vMol \rangle_{2 \rho \rightarrow 1 \rho}^{\rm tree} \Bigg\{ 2 Y_\rho^{\rm eq} \bigg[ \Delta_\rho \left( 1- \frac{Y_2^{\rm eq}}{Y_1^{\rm eq}}\right) + \frac{\beta- \bar{\beta}}{2} \left( 1 + \frac{Y_2^{\rm eq}}{Y_1^{\rm eq}} \right) \!\bigg] \nonumber \\
    &-(\epsilon_{2 \rho \to 1 \rho} - \epsilon_{\bar{2} \rho \to \bar{1} \rho})\Bigg[ \frac{\gamma + \bar{\gamma}}{2} \left(1+\frac{Y_2^{\rm eq}}{Y_1^{\rm eq}} \right) - \frac{\beta+ \bar{\beta}}{2} \left(1 - \frac{Y_2^{\rm eq}}{Y_1^{\rm eq}} \right) \! \Bigg]  Y_\rho^{\rm eq}\nonumber \\
    &+(\epsilon_{2 \rho \to 1 \rho} + \epsilon_{\bar{2} \rho \to \bar{1} \rho})\Bigg[\Delta_\rho \left(1 + \frac{Y_2^{\rm eq}}{Y_1^{\rm eq}} \right) + \frac{\beta - \bar{\beta}}{2} \left(1  - \frac{Y_2^{\rm eq}}{Y_1^{\rm eq}}\right) \Bigg]  \frac{\Delta_\rho}{2}  \Bigg\}.
\end{align}
\end{subequations}
\end{widetext}
\renewcommand{\arraystretch}{2}
\begin{table*}
\resizebox{\textwidth}{!}{
\begin{tabular}{|c| c| c| c | c |c|}
\hline
\Large{\bf Process}  
& \Large{\bf Channel (mediator)} 
& \Large{$\bm{16 \pi \cdot \sigma_{\rm tree}}$} 
& \Large{\bf Sommerfeld} 
& \large{$\bm{16 \pi  \cdot \varepsilon}$}  
\\ [0.5ex] 
\hline
\multicolumn{5}{|c|}
{\Large{In the following:~~~ 
\phantom{\bigg(}
$\mathbb{Y} \equiv 16 \cdot  
\left|\y_{11} \y_{12}^* + \y_{12} \y_{22}^* \right|^2,
\quad
\y_{ij} \equiv \frac{1}{2}(y_{ij}^{\rm L} + y_{ij}^{\rm R}),  
\quad
\w_{ij} \equiv \frac{1}{2}(y_{ij}^{\rm L} - y_{ij}^{\rm R}), 
\qquad {\rm for} \quad i,j \in \{1,2\}$
}} 
\\ \hline \hline
\multicolumn{5}{|c|}
{\phantom{\bigg(} \Large{$\bm{\qBL=-3^{}}$} \phantom{\bigg)}} 
\\ \hline
\large{$X_2 \rho \rightarrow X_1 \rho$} 
& \large{u$-$ ($X_i$)}  
& \large{$ s/(s-m_2^2)^2 \cdot \mathbb{Y} 
$} 
& $\tikzxmark$  
& \large{
$\phantom{\Bigg[} 
\dfrac{ 8 \cdot \Im{} \left(\y_{12}^2 \y_{11}^* \y_{22}^*\right)}
{\mathbb{Y}} \bigg( 8 \cdot (|\y_{11}|^2 + |\y_{12}|^2) + \gBL^2 \cdot \ln{2}  \bigg)
$}  

\\ \hline \hline
\multicolumn{5}{|c|}
{\phantom{\bigg(} \Large{$\bm{\qBL=-2^{}}$} \phantom{\bigg)}} 
\\ \hline
\large{$X_2 X_2 \rightarrow X_1 X_1$} 
&  \large{s$-$ ($\rho$)} 
& \large{$  \phantom{\Bigg(}(16 \cdot m_2^2)^{-1} \cdot |\w_{11}|^2 \cdot |\w_{22}|^2$} 
& \large{$ S_0(-\zeta_{\rm{\mathsmaller{B-L}}}) \cdot S_0(-\zeta_{\rm{\mathsmaller{B-L}}}^\prime ) $} 
& \large{$\dfrac{4 \cdot  \Im{} (\y_{11}^* \y_{12}  \,\w_{12} \w_{22}^*)
}{ |\w_{22}|^2} $
}   
\\ \hline
\large{$X_2 X_1 \rightarrow X_1 X_1$} 
& \large{s$-$ ($\rho$)} 
& \large{$  \phantom{\Bigg(}( 16 \cdot m_2^2)^{-1} \cdot |\w_{12}|^2 \cdot |\w_{11}|^2$} 
& \large{$ S_0(-\zeta_{\rm{\mathsmaller{B-L}}}) \cdot S_0(-\zeta_{\rm{\mathsmaller{B-L}}}^\prime) $} 
& \large{
$\dfrac{4 \cdot  \Im{} (\y_{11} \y_{12}  \,\w_{11}^* \w_{12}^*)
}{ |\w_{12}|^2} $
}   
\\  \hline
\large{$X_2 X_1 \rightarrow X_2 X_2$} 
& \large{s$-$ ($\rho$)} 
& \large{$  \phantom{\Bigg(}  (16 \cdot m_2^2)^{-1}  \cdot |\w_{12}|^2 \cdot |\w_{22}|^2$} &  \large{$ S_0(-\zeta_{\rm{\mathsmaller{B-L}}}) \cdot S_0(-\zeta_{\rm{\mathsmaller{B-L}}}^\prime) $}  
& \large{
$\dfrac{4 \cdot  \Im{} (\y_{11} \y_{12}^*  \,\w_{12}^2-* \w_{22})
}{ |\w_{22}|^2} + \dfrac{4 \cdot  \Im{} (\y_{11} \y_{12}  \,\w_{11}^* \w_{12}^*)
}{ |\w_{12}|^2}  $
}  
\\ \hline
\large{$X_1 X_1 \rightarrow \rho V_\mu$} & \large{s$-$ ($\rho$), t$-$ \& u$-$  ($X_1$)} & \large{$  \phantom{\Bigg(} (\vrel \, m_1^2 )^{-1} \cdot  \gBL^2  \cdot |\y_{11}|^2$} & \large{$S_0(-\zeta_{\rm{\mathsmaller{B-L}}})$} & $\tikzxmark$ 
\\ \hline
\large{$X_2 X_1 \rightarrow \rho V_\mu$} & \large{s$-$ ($\rho$), t$-$ ($X_1$), u$-$ ($X_2$)} & \large{$ \phantom{\Bigg(} (\vrel \,m_2^2)^{-1} \cdot  \gBL^2 \cdot |\y_{12}|^2$} &  \large{$S_0(-\zeta_{\rm{\mathsmaller{B-L}}})$} & $\tikzxmark$   
\\ \hline
\large{$X_2 X_2 \rightarrow \rho V_\mu$}& \large{s$-$ ($\rho$), t$-$ \& u$-$  ($X_2$)}  &  \large{$ \phantom{\Bigg(} (\vrel \,m_2^2)^{-1} \cdot  \gBL^2 \cdot |\y_{22}|^2$}   & \large{$S_0(-\zeta_{\rm{\mathsmaller{B-L}}})$}  & $\tikzxmark$  
\\ \hline \hline
\multicolumn{5}{|c|}{ \phantom{\bigg(} \Large{$\bm{\qBL=-1^{}}$} \phantom{\bigg)}} \\
\hline
\large{$\bar{X}_2 \rho \rightarrow \bar{X}_1 \rho$} &\large{s$-$ ($X_i$)} 
& \large{$s/(s-m_2^2)^2 \cdot \mathbb{Y}$} 
& $\tikzxmark$ 
&  \large{
$\phantom{\Bigg[} 
\dfrac{8 \cdot \Im{} \left(\y_{12}^{*2} \y_{11} \y_{22} \right)}
{\mathbb{Y}} \bigg( 8 \cdot (|\y_{11}|^2 + |\y_{22}|^2) + \gBL^2 \cdot \ln{2} \bigg)$
}
\\ \hline
\large{$X_1 V_\mu \rightarrow \bar{X}_1 \rho$} 
&  \large{s$-$ ($X_1$),  t$-$ ($\rho$),   u$-$ ($X_1$)} 
& \large{$ \phantom{\Bigg(} 2\sqrt{s}/(3\cdot m_1 (s-m_1^2)) \cdot g^2_{\rm B-L} \cdot |\y_{11}|^2 $} 
& $\tikzxmark$& $\tikzxmark$  
\\ \hline
\large{$X_1 V_\mu \rightarrow \bar{X}_2 \rho$} 
&  \large{s$-$ ($X_1$),  t$-$ ($\rho$),  u$-$ ($X_2$)}
& \large{$ \phantom{\Bigg(} 2\sqrt{s}/(3\cdot m_2 (s-m_2^2)) \cdot g^2_{\rm B-L} \cdot |\y_{12}|^2 $} &  $\tikzxmark$ 
& $\tikzxmark$ 
\\ \hline
\large{$X_2 V_\mu \rightarrow \bar{X}_1 \rho$} 
&  \large{s$-$ ($X_2$), t$-$ ($\rho$),  u$-$ ($X_1$)}
& \large{$ \phantom{\Bigg(} 2\sqrt{s}/(3\cdot m_2 (s-m_2^2)) \cdot g^2_{\rm B-L}  \cdot |\y_{12}|^2 $} 
& $\tikzxmark$ 
&  $\tikzxmark$  
\\ \hline 
\large{$X_2 V_\mu \rightarrow \bar{X}_2 \rho$} &  \large{s$-$ ($X_2$), t$-$ ($\rho$),  u$-$ ($X_2$)} &  \large{$ \phantom{\Bigg(} 2\sqrt{s}/(3\cdot m_2 (s-m_2^2))  \cdot g^2_{\rm B-L}   \cdot |\y_{22}|^2 $} & $\tikzxmark$ &  $\tikzxmark$ 
\\ \hline \hline
\multicolumn{5}{|c|}{\phantom{\bigg(} 
\Large{$\bm{\qBL=0}$} 
\phantom{\bigg)} } 
\\ \hline
\large{$ \rho \rho^* \rightarrow V_\mu V_\mu$} 
&  \large{t$-$ \& u$-$ $(\rho)$, 4-point vertex} 
& \large{$ \phantom{\Bigg(} 64/s \cdot g^4_{\rm B-L}$}
&  $\tikzxmark$ & $\tikzxmark$ 
\\ \hline 
\large{$X_1 \bar{X}_1 \rightarrow V_\mu V_\mu$} & \large{t$-$ \& u$-$ ($X_1$)} & \large{$\phantom{\Bigg(}(\vrel \, m_1^2 )^{-1} \cdot \gBL^4$} 
& \large{$\chk$}
& $\tikzxmark$  
\\ \hline
\large{$ X_2 \bar{X}_2 \rightarrow V_\mu V_\mu$} & \large{t$-$ \& u$-$ ($X_2$)} & \large{$\phantom{\Bigg(}( \vrel\, m_2^2 )^{-1} \cdot \gBL^4$} 
& \large{$\chk$}
& $\tikzxmark$ 
\\ 
\hline
\large{$ X_1 \bar{X}_2 \rightarrow V_\mu V_\mu$} 
& \large{$-$} 
& \large{$0$} 
& \large{$\chk$} 
& $\tikzxmark$ 
\\ 
\hline 
\large{$ X_2 \bar{X}_1 \rightarrow V_\mu V_\mu$} 
& \large{$-$} 
& \large{$0$} 
& \large{$\chk$} 
& $\tikzxmark$ 
\\ 
\hline 
\large{$ X_1 \bar{X}_1 \rightarrow \rho \rho^*$} 
& \large{s$-$ ($V_\mu$), t$-$ $(X_i)$}
& \large{\makecell{ \phantom{\Bigg[}
$(4 \cdot  \vrel\, m_1^2)^{-1}\cdot  \bigg[ \bigg( 2 \gBL^2 + \displaystyle\sum_{i=1}^2 (|\y_{1i}|^2 + |\w_{1i}|^2) \bigg)^2 + 4 \cdot 
\bigg( \displaystyle\sum_{i=1}^2 \Im {}(\w_{1i} \y_{1i}^{*} )\bigg)^2
\bigg] $  \phantom{\Bigg]}}}  
& \large{$\chk$}
& $\tikzxmark$ 
\\ \hline
\large{$ X_1 \bar{X}_2 \rightarrow \rho \rho^*$} &  \large{t$-$ $(X_i)$} & 
\large{\makecell{ \phantom{\Bigg[}
$(4 \cdot  \vrel \, m_2^2)^{-1} \cdot  \bigg[ \bigg( \displaystyle\sum_{i=1}^2  (\y_{1i} \y_{i2}^{*} + \w_{1i} \w_{i2}^{*}) \bigg)^2 -  \bigg( \displaystyle\sum_{i=1}^2  (\y_{1i} \y_{i2}^{*} - \w_{1i} \w_{i2}^{*}) \bigg)^2 \bigg]$ \phantom{\Bigg]}
}} 
& \large{$\chk$} 
& \large{$\chk$}
\\ \hline 
\large{$ X_2 \bar{X}_2 \rightarrow \rho \rho^*$} 
& \large{s$-$ ($V_\mu$), t$-$ $(X_i)$}  
& \large{\makecell{ \phantom{\Bigg[}
$(4 \cdot  \vrel\, m_2^2)^{-1}\cdot  \bigg[ 
\bigg(2 \gBL^2 + \displaystyle\sum_{i=1}^2 (|\y_{i2}|^2 + |\w_{i2}|^2) \bigg)^2 
+ 4 \cdot  
\bigg( \displaystyle\sum_{i=1}^2 \Im {}(\w_{i2} \y_{i2}^{*} ) \bigg)^2
\bigg]$ \phantom{\Bigg]} }}  
& \large{$\chk$} 
& $\tikzxmark$ 
\\ \hline
\end{tabular}
}
\caption{Tree-level cross-sections, $\sigma_{\rm tree}$, for 2-to-2 scatterings, grouped according to their total charge $\qBL=\{-3, -2, -1, 0 \}$. We do not include here the elastic-like processes, $X_i \bar{X}_j \rightarrow X_{i^\prime} \bar{X}_{j^\prime}$, which are discussed in a companion paper~\cite{Flores:2025b}. Note that $\sigma_{\rm tree}$ is summed over the final spin states and averaged over the initial spin states. In the case of identical final state we include the $1/2$ factor. For convenience, the cross-sections for processes involving at least one relativistic particle in the initial state are expressed as a function of the Mandelstam variable $s$. The Sommerfeld suppression factors for the $X_i X_j$ interactions are defined in \cref{eq:Sommerfeld}.
The above results neglect the $X$ mass difference, i.e., $\Delta \mX \rightarrow0$, as well as the masses of the force mediators.}
\label{tab:CrossSections}
\end{table*}

\end{document}